\documentclass[aps,preprint,onecolumn,12pt,nofootinbib]{revtex4}
\usepackage{hyperref}
\usepackage{amsfonts}
\usepackage{multirow}
\usepackage{graphicx}
\usepackage{placeins}
\usepackage[utf8]{inputenc}
\usepackage{amsmath}

\usepackage{amssymb}
\usepackage{bbm}
\usepackage{tensor}
\usepackage{todonotes}
\usepackage{graphicx}%
\usepackage{bigints}
\usepackage{subfigure}
\usepackage{epsf}
\usepackage{bm}
\usepackage{amsmath}
\usepackage{amsfonts}
\usepackage{amssymb}
\usepackage{tabularx}
\usepackage{color}%
\providecommand{\U}[1]{\protect\rule{.1in}{.1in}}

\newcommand{\ie}{\begin{equation}}
\newcommand{\fe}{\end{equation}}

\newcommand{\mincir}{\raise
-3.truept\hbox{\rlap{\hbox{$\sim$}}\raise4.truept\hbox{$<$}\ }}
\newcommand{\magcir}{\raise
-3.truept\hbox{\rlap{\hbox{$\sim$}}\raise4.truept\hbox{$>$}\ }}

\providecommand{\U}[1]{\protect\rule{.1in}{.1in}}

\usepackage{tikz,xcolor,hyperref}

\definecolor{lime}{HTML}{A6CE39}
\DeclareRobustCommand{\orcidicon}{%
	\begin{tikzpicture}
	\draw[lime, fill=lime] (0,0) 
	circle [radius=0.16] 
	node[white] {{\fontfamily{qag}\selectfont \tiny ID}};
	\draw[white, fill=white] (-0.0625,0.095) 
	circle [radius=0.007];
	\end{tikzpicture}
	\hspace{-2mm}
}

\foreach \x in {A, ..., Z}{%
	\expandafter\xdef\csname orcid\x\endcsname{\noexpand\href{https://orcid.org/\csname orcidauthor\x\endcsname}{\noexpand\orcidicon}}
}

\begin{document}

\title{\Large{Imprints of Non--commutativity on Charged Black Holes}}


\author{N. Heidari\orcidA{}}
\email{heidari.n@gmail.com}

\affiliation{Center for Theoretical Physics, Khazar University, 41 Mehseti Street, Baku, AZ-1096, Azerbaijan.}



	\begin{abstract}
    This work presents a comprehensive investigation of the gravitational phenomena that correspond to a non--commutative charged black hole, by incorporating non--commutative geometry through a $(r,\theta)$ Moyal twist. We derive the deformed metric up to the second order of non--commutative parameter $\Theta$, utilizing the Seiberg--Witten map for Reissner--Nordstr\"{o}m black hole. We explore how non--commutativity modifies key thermodynamic properties, such as the Hawking temperature and heat capacity, and the existence of a remnant mass at the final stage of the evaporation. Additionally, the study of Hawking radiation for bosonic and fermionic particles is discussed. Applying a perturbative method, scalar quasinormal modes are analyzed numerically. Furthermore, null geodesics and photon sphere stability are explored via curvature and topological methods. The shadow radius and deflection angle are computed to understand observational signatures. Lensing observables are compared to Event Horizon Telescope (EHT) observations to provide probable constraints on the non--commutativity parameter. This study bridges theoretical predictions with astrophysical observations, offering insights into quantum gravity effects on black hole physics.
	\end{abstract}
\maketitle

\tableofcontents

\section{Introduction}

In the framework of general relativity (GR), black holes arise as fundamental solutions to Einstein’s field equations. However, the classical description of these objects becomes inadequate at the Planck scale, where quantum gravitational effects are expected to dominate. A common aspect of several quantum gravity candidates, including string theory and matrix models, is the emergence of a \textit{minimal length scale}.
Although many approaches, such as string theory and loop quantum gravity, have made progress in describing quantum aspects of spacetime, there is currently no conclusive theory of quantum gravity, which motivates the study of non--commutative (NC) spacetimes~\cite{Snyder1947, SeibergWitten1999, Szabo2003,connes1998noncommutative,ardalan1999noncommutative} as a candidate with a geometrical approach. NC geometry introduces a deformation of the spacetime manifold through noncommuting coordinates, typically written as
\begin{equation}\label{NCbase}
[x^\mu, x^\nu] = i \Theta^{\mu\nu},
\end{equation}
where $\Theta^{\mu\nu}$ is a constant antisymmetric matrix that characterizes the spacetime deformation. This deformation yields a fundamental length scale below which the classical concept of spacetime continuity ceases to apply. Such ideas emerge naturally from string theory, matrix models, making NC geometry a compelling framework for encoding Planck--scale corrections~\cite{DouglasNekrasov2001, Connes1994, Nicolini2009}. 

Various approaches have been presented to include NC adjustments into gravity, including the smearing of matter sources~\cite{Nicolini2006, Spallucci2009,nozari2008hawking,araujo2024geodesics}, use of the Moyal star product~\cite{Smailagic2003}, and more systematically, the application of the Seiberg--Witten (SW) map~\cite{jurvco2001construction, Chaichian2002, Aschieri2005,banerjee2007lie,chaichian2008corrections}. The SW map provides the expansion of NC fields in terms of their commutative counterparts, ensuring gauge covariance and offering a controlled framework for embedding NC effects into gravitational systems. This approach has been vastly applied in the investigation of black hole physics in NC spacetime in recent years ~\cite{chaichian2008black,mukherjee2008deformed,linares2020thermodynamical,AraujoFilho:2025jcu,heidari2023gravitational,heidari2025non,bevzanic2025noncommutative,ciric2018noncommutative,herceg2025noncommutative,AraujoFilho:2025rvn,chamseddine2023noncommutativity}. NC black holes have been shown to regularize curvature singularities~\cite{Nicolini2006, Spallucci2009}, modify thermodynamic characteristics including the Hawking temperature and entropy~\cite{nozari2008hawking,Banerjee2011,araujo2023thermodynamics,banerjee2008noncommutative,mehdipour2010hawking}, and yield potentially observable signatures in gravitational wave signals and shadow profiles~\cite{Zhao2024, Lin2019,heidari2024exploring,campos2022quasinormal,wei2015shadow}.

In this work, we follow the NC modifications of black hole solutions introduced in Ref.~\cite{juric2025constructing} and investigate the imprint of the NC framework on a charged black hole. We introduce a Moyal--type twist in the $(r, \theta)$ coordinate sector and construct a consistent NC-deformed metric by applying the Seiberg--Witten map to the vierbein fields. Second-order corrections to the spin connection are included to ensure consistency up to $\mathcal{O}(\Theta^2)$, which captures subtle quantum corrections.

Our analysis spans several interconnected aspects of NC black hole physics. We start by examining the thermodynamic characteristics of the NC-Reissnerr Nordstro\"{o}m black hole, including the modification of the Hawking temperature. We also explore quantum tunneling of both bosonic and fermionic particle modes, revealing deformation--induced shifts in the radiation spectrum. A major challenge in this context is the computation of quasinormal modes (QNMs), which express the black hole’s response to external perturbations. In NC spacetimes, the equation for scalar perturbations becomes highly nontrivial due to the complex form of the metric functions. We address this by deriving a Schr\"{o}dinger--like master equation through a novel numerical solution via perturbation method ~\cite{Zhao2024, Lin2019,heidari2024exploring}  and compute QNMs, using the WKB approximation.
Moreover, we investigate the structure of the \textit{photon sphere}, which governs key observational features such as black hole shadows. The stability of the photonic sphere is explored via both curvature and topological approaches. We compute the shadow radius and the deflection angles in the NC geometry and compare our predictions of the lensing observables with Event Horizon Telescope (EHT) observations of $M87^*$ and $Sgr A^*$. These comparisons allow us to check the probable constraints on the NC deformation parameter $\Theta$, thereby assessing the observational viability of NC gravity models.

In summary, this work presents a comprehensive investigation of a charged black hole in a non--commutative spacetime. By connecting theoretical predictions with observational data, we aim to bridge the gap between quantum gravity models and astrophysical measurements, providing potential empirical opportunities to explore the physics of spacetime at the smallest scale.


\section{Non--commutative charged black hole}
The non--commutative (NC) spacetime is introduced through the fundamental commutation relation in Eq. \eqref{NCbase}. Following the methodology outlined in Ref.~\cite{juric2025constructing}, we derive the non--commutative corrections to the Reissner–Nordstr\"{o}m metric, with further details available therein. The Moyal twist element is defined as
\begin{equation}
    \mathcal{F} = \exp\left(-\frac{i}{2}\Theta^{\mu\nu}\partial_\mu \otimes \partial_\nu\right),
\end{equation}
which induces the star product between functions
\begin{equation}
    f \star g = m \footnotemark \left(\mathcal{F}^{-1} f \otimes g\right).
\end{equation}
\footnotetext{m is the commutative multiplication map
$m (a\bigotimes b) = a \cdot  b$.}
The gauge transformation in the NC framework becomes
\begin{equation}
    \hat{\delta}_{\hat{\lambda}} \hat{\omega}_\mu = \partial_\mu \hat{\lambda} + i[\hat{\lambda}, \hat{\omega}_\mu]_\star,
\end{equation}
where the star commutator is
\begin{equation}
    [\hat{\lambda}, \hat{\omega}_\mu]_\star = \hat{\lambda} \star \hat{\omega}_\mu - \hat{\omega}_\mu \star \hat{\lambda}.
\end{equation}

Following Seiberg--Witten map, the next expansion for the deformed tetrad fields is provided

\begin{equation}
    \hat{e}_\mu^a(r, \Theta) = e_\mu^a(x) - i\Theta^{\nu\rho}e_{\mu\nu\rho}^a(x) + \Theta^{\nu\rho}\Theta^{\lambda\tau}e_{\mu\nu\rho\lambda\tau}^a(x) + \mathcal{O}(\Theta^3),
\end{equation}

with
\begin{align}
e_{\mu \nu \rho }^{a}
&= \frac{1}{4}\Bigl[
   \omega_{\nu}^{a\,c}\,\partial_{\rho}\,e_{\mu}^{d}
   \;+\;\bigl(\partial_{\rho}\,\omega_{\mu}^{a\,c} + R_{\rho \mu}^{a\,c}\bigr)\,e_{\nu}^{d}
\Bigr]\;\eta_{c\,d}, 
\label{3.11} 
\\
e_{\mu \nu \rho \lambda \tau }^{a}
&= \frac{1}{16}\Bigl[
   2\,\Bigl\{R_{\tau \nu},\,R_{\mu \rho}\Bigr\}^{ab}\,e_{\lambda}^{c}
   \;-\;\omega_{\lambda}^{a\,b}\,\Bigl(D_{\rho}\,R_{\tau \mu}^{c\,d}
     \;+\;\partial_{\rho}\,R_{\tau \mu}^{c\,d}\Bigr)\,e_{\nu}^{m}\,\eta_{d\,m}
   \nonumber \\
&\quad
-\,\Bigl\{\omega_{\nu},\,\bigl(D_{\rho}\,R_{\tau \mu}
     + \partial_{\rho}\,R_{\tau \mu}\bigr)\Bigr\}^{ab}\,e_{\lambda}^{c}
     \;-\;\partial_{\tau}\,\Bigl\{\omega_{\nu},\,\bigl(\partial_{\rho}\,\omega_{\mu}
     + R_{\rho \mu}\bigr)\Bigr\}^{a\,b}\,e_{\lambda}^{c}
   \nonumber \\
&\quad
-\,\omega_{\lambda}^{a\,b}\,\partial_{\tau}\Bigl(
       \omega_{\nu}^{c\,d}\,\partial_{\rho}\,e_{\mu}^{m}
       + \bigl(\partial_{\rho}\,\omega_{\mu}^{c\,d}
         + R_{\rho \mu}^{c\,d}\bigr)\,e_{\nu}^{m}
     \Bigr)\,\eta_{d\,m}
     \nonumber \\
&\quad
   \;+\;2\,\partial_{\nu}\,\omega_{\lambda}^{a\,b}\,
        \partial_{\rho}\partial_{\tau}\,e_{\mu}^{c}
   -\,2\,\partial_{\rho}\Bigl(\partial_{\tau}\,\omega_{\mu}^{a\,b}
     + R_{\tau \mu}^{a\,b}\Bigr)\,\partial_{\nu}\,e_{\lambda}^{c}
   \;-\;\Bigl\{\omega_{\nu},\,\bigl(\partial_{\rho}\,\omega_{\lambda}
     + R_{\rho \lambda}\bigr)\Bigr\}^{a\,b}\,\partial_{\tau}\,e_{\mu}^{c}
\nonumber \\
&\quad
   \;-\,\Bigl(\partial_{\tau}\,\omega_{\mu}^{a\,b}
     + R_{\tau \mu}^{a\,b}\Bigr)\,\Bigl(
       \omega_{\nu}^{c\,d}\,\partial_{\rho}\,e_{\lambda}^{m}
       + \bigl(\partial_{\rho}\,\omega_{\lambda}^{c\,d}
         + R_{\rho \lambda}^{c\,d}\bigr)\,e_{\nu}^{m}\,\eta_{d\,m}
     \Bigr)
\Bigr]\;\eta_{b\,c}
\nonumber \\
&\quad
\; - \frac{1}{16}\,\omega_{\lambda}^{a\,c}\,\omega_{\nu}^{d\,b}\,e_{\rho}^{f}\,
   R_{\tau \mu}^{g\,m}\,\eta_{c\,d}\,\eta_{f\,g}\,\eta_{b\,m}\,,
\label{NCtetrad}
\end{align}
and the spin connection fields are expressed as 
\begin{equation} \label{spincorr}
    \hat \omega _\mu ^{a b } (r, \Theta)= \omega _\mu ^{a b} (x)-i \Theta ^{\nu \rho} \omega _{\mu \nu \rho} ^{a b} (x) +\Theta ^{\nu \rho} \Theta ^{\lambda \tau} \omega _{\mu \nu \rho \lambda \tau} ^{a b} (x)+\mathcal{O}(\Theta ^3)
\end{equation}
with
\begin{equation}
	\omega_{\mu \nu \rho}^{ab} (x) = \frac{1}{4}\left\{{\omega}_\nu , \partial_\rho {\omega}_\mu + {R}_{\rho\mu}\right\}^{ab}\;,
\end{equation}
and
\begin{equation}
	\begin{split}
		\omega_{\mu\nu\rho\lambda\tau}^{ab} = &\frac{1}{16}\biggl[-\left\{\left\{\omega_\lambda,\left(\partial_\tau\omega_\nu + R_{\tau\nu}\right)\right\},\left(\partial_\rho\omega_\mu + R_{\rho\mu}\right)\right\}^{ab} \\
  & - \left\{\omega_\nu, \partial_\rho \left\{\omega_\lambda,\left(\partial_\tau\omega_\mu + R_{\tau\mu}\right)\right\} \right\}^{ab}  +2 \left[\partial_\lambda\omega_\nu, \partial_\tau\left(\partial_\rho\omega_\mu + R_{\rho\mu}\right)\right]^{ab}  \\
  & + \left\{\omega_\nu,2\left\{R_{\rho\lambda},R_{\mu\tau}\right\}\right\}^{ab} - \left\{\omega_\nu,\left\{\omega_\lambda, D_\tau R_{\rho\mu} + \partial_\tau R_{\rho \mu}\right\}\right\}^{ab}   \biggr].
	\end{split}
\label{22b}
\end{equation}
Utilizing the NC tetrad fields $\hat{e}^a_\mu$ introduced  in~\cite{chaichian2008corrections}, deformed NC metric is proposed as
\begin{equation}\label{metdef}
    \hat{g}_{\mu\nu} = \frac{1}{2}\eta_{ab}\left(\hat{e}^a_\mu \star \hat{e}^{b\dagger}_\nu + \hat{e}^{b}_\nu\star \hat{e}^{a\dagger}_\mu\right)
\end{equation}
Now, the NC gauge transformation obeying counterpart $\hat{e}^a_\mu$ results in the construction of the NC black hole metric.

As mentioned before, this approach has been extensively employed in the literature to examine the phenomenology of spacetime within the NC spacetime ~\cite{chaichian2008black,mukherjee2008deformed,linares2020thermodynamical,heidari2023gravitational,heidari2025non,bevzanic2025noncommutative,touati2024non}. However,~\cite{juric2025constructing} proposed that a correction term be applied to \eqref{NCtetrad}, which makes the whole prior investigation revisited.\\

We present the metric for the Reissner--Nordstr\"{o}m black hole in the $\partial_r \wedge \partial_\theta $ Moyal twist. We correct prior errors--specifically addressing the overlooked term in Eq. \eqref{NCtetrad}--and introduce new, previously unexplored analyses for this significant NC metric.

The twist in the $(r,\theta)$ plane is non--Killing and is characterized by the tensor $\Theta^{\mu\nu}$.
\begin{equation}
\Theta ^{\mu \nu }=\left(
\begin{array}{cccc}
0 & 0 & 0 & 0 \\
0 & 0 & \Theta  & 0 \\
0 & -\Theta & 0 & 0 \\
0 & 0 & 0 & 0
\end{array}
\right) ,\quad \mu ,\nu =0,1,2,3.  \label{4drtheta}
\end{equation}
With this choice, the coordinates satisfy the commutation relation
\begin{equation}
    [r\stackrel{\star}{,} \theta] = i \, \Theta. 
\end{equation}

Although this twist has been investigated in the literatures ~\cite{chaichian2008corrections,mukherjee2008deformed,touati2024quantum}, we revisit it here for Reissner--Nordstr\"{o}m considering the newly identified terms in \eqref{NCtetrad}, with unexplored aspects of a charged black hole in this framework. 
Based upon this twist, the non-zero elements of the tetrad $e^a_\mu$ include as following
\begin{eqnarray}
\hat{e}_{0}^{0}&=&A+\frac{1}{4}\,\left( 2\,r\,{A}^{\prime
}{}^{3}+5\,r\,A\,{A}^{\prime }\,{A}^{\prime \prime
}+r\,A^{2}\,{A}^{\prime \prime \prime }+2\,A\, {A}^{\prime
}{}^{2}+A^{2}\,{A}^{\prime \prime }\right) \Theta ^{2}+O(\Theta {^{3}})\,,  \label{4drthetab}\\
\hat{e}_{1}^{1}&=&\frac{1}{A}+\frac{{A}^{\prime \prime}}{4}\,\Theta
^{2}+O(\Theta {^{3}}),  \cr
\hat{e}_{2}^{1}&=&-\frac{i}{4}\left( A+2\,r\,{A}^{\prime }\right)
\,\Theta +O(\Theta {^{3}}) ,  \cr
\hat{e}_{2}^{2}&=&r+\frac{1}{4}\,\left( 2A\,{A}^{\prime
}+3\,r\,{A}^{\prime }{}^{2}+3\,r\,A\,{A}^{\prime \prime }\right)
\Theta ^{2}+O(\Theta {^{3}}) ,  \cr
\hat{e}_{3}^{3}&=&r\sin \theta -\frac{i}{4}\left( \cos \theta
\right) \Theta +\frac{1}{4}\,\left( 2r\,{A}^{\prime }{}^{2}+r A
{A}^{\prime \prime }+2A{A}^{\prime }-\frac{{A}^{\prime }}{A}\right)
\sin \theta \,\Theta ^{2}+O(\Theta {^{3}}), \nonumber
\end{eqnarray}
here ${A}^{\prime },\,{A}^{\prime \prime },\,{A}^{\prime \prime
\prime }$ represent the first, second, and third derivatives of $A(r)$,
respectively. The deformed metric $\hat g _{\mu \nu} (r, \Theta)$ is obtained by using the definition in Eq. \eqref{metdef}. The resulting metric is diagonal, and the non--zero components, up to the second order of $\Theta$, are
\begin{eqnarray}
\hat{g}_{tt}\left( r,\Theta \right) &=&-A^{2}-\frac{1}{2}\,\left(
2\,r\,A\,{ A}^{\prime }{}^{3}+r\,A^{3}\,{A}^{\prime \prime \prime
}+A^{3}\,{A}^{\prime \prime }+2\,A^{2}\,{A}^{\prime
}{}^{2}+5\,r\,A^{2}\,{A}^{\prime }\,{A} ^{\prime  \prime }\right)
{\Theta }^{2},\cr
\hat{g}_{rr}\left( r,\Theta \right)
&=&\frac{1}{A^{2}}+\frac{1}{2}\,\frac{{ A}^{\prime \prime
}}{A}\,{\Theta }^{2}, \label{NCrtheta1}\\
\hat{g}_{\theta\theta}\left( r,\Theta \right) &=&r^{2}+\frac{1}{16}\,\left(
A^{2}+20\,r\,A\,{A}^{\prime }+28\,r^{2}\,{A}^{\prime
}{}^{2}+24\,r^{2}A\,{A} ^{\prime \prime }\right) {\Theta }^{2}, \cr
\hat{g}_{\varphi\varphi}\left( r,\Theta \right) &=&r^{2}\,\sin ^{2}\theta
\cr
&+&\frac{1 }{16}\left[4\,\left(4\,r\,A\,{A}^{\prime
}-\,2r\frac{{A}^{\prime }}{A} +2\,r^{2}\,A\,{A}^{\prime \prime
}+4\,r^{2}\,{A}^{\prime }{}^{2} +1\right) \sin ^{2}\theta +5\cos
^{2}\theta \right] {\Theta }^{2}. \nonumber
\end{eqnarray}

The NC-corrected Reisser--Nordstr\"{o}m metric components under the $(r, \theta)$ twist are obtained by substituting the Reissner--Nordstr\"{o}m black hole function $A(r)=\sqrt{1-\frac{2M}{r}+\frac{Q^2}{r^2}}$ into Eq. \eqref{NCrtheta1},

\begin{subequations}\label{metric1}
\begin{align}
g^{\scriptscriptstyle {(\Theta},\scriptscriptstyle{Q)}}_{tt}=&
{-}(1-\frac{2M}{r}+\frac{Q^2}{r^2}
)+\frac{ \left(3 Q^2 r (3 r-10 M)+M r^2 (11 M-4 r)+14 Q^4\right)\Theta ^2}{2 r^6},\\ \label{metric2}
g^{\scriptscriptstyle {(\Theta},\scriptscriptstyle{Q)}}_{rr}=&(1-\frac{2M}{r}+\frac{Q^2}{r^2}
)^{-1}+\frac{\Theta ^2 \left(3 Q^2 r (r-2 M)+M r^2 (3 M-2 r)+2 Q^4\right)}{2 r^2 \left(r (r-2 M)+Q^2\right)^2},\\ 
g^{\scriptscriptstyle {(\Theta},\scriptscriptstyle{Q)}}_{\theta\theta}=&r^2+\frac{\Theta ^2 \left(r^2 \left(64 M^2-32 M r+r^2\right)+18 Q^2 r (3 r-8 M)+57 Q^4\right)}{16 r^2 \left(r (r-2 M)+Q^2\right)},\\ 
g^{\scriptscriptstyle {(\Theta},\scriptscriptstyle{Q)}}_{\phi \phi}=&r^2\sin^2\theta\\ \nonumber
+&\frac{\Theta ^2}{16}\left(5 \cos ^2\theta+\frac{4 \sin ^2\theta \left(r^2 \left(2 M^2-4 M r+r^2\right)+Q^2 r (5 r-8 M)+4 Q^4\right)}{r^2 \left(r (r-2 M)+Q^2\right)}\right).
\end{align}
\end{subequations}


\section{Thermodynamics}

We start our analysis of the above spacetime by turning our attention to the thermodynamic aspects of the Reisser--Nordstr\"{o}m black hole within the non--commutative gauge theory. Specifically, we examine the behavior of the Hawking temperature, entropy, and heat capacity. We will express all thermodynamic properties as functions of the event horizon radius $r_{h}$, as it is a standard procedure in the literature. In particular, special emphasis will be placed on the Hawking temperature, which will also be studied as a function of the black hole mass $M$ to assess the possible existence of a remnant mass.

\subsection{Hawking temperature}

This section will be devoted to investigating the Hawking temperature. Using the surface gravity procedure which reads \cite{araujo2024dark,araujo2024charged,araujo2023thermodynamical}
\ie
\begin{split}
\label{htemp}
T_{H} & =  \frac{1}{4 \pi} \frac{1}{\sqrt{g^{\scriptscriptstyle {(\Theta},\scriptscriptstyle{Q)}}_{tt}(r) \, g^{\scriptscriptstyle {(\Theta},\scriptscriptstyle{Q)}}_{rr}(r)}} \left. \frac{\mathrm{d}g^{\scriptscriptstyle {(\Theta},\scriptscriptstyle{Q)}}_{tt}(r)}{\mathrm{d}r}  \right|_{r = r_{h}} \\
& = \frac{2 r_{h}^4 \left(M r_{h}-Q^2\right)+\Theta ^2 \left(3 Q^2 r_{h} (6 r_{h}-25 M)+2 M r_{h}^2 (11 M-3 r_{h})+42 Q^4\right)}{2 \pi  r_{h}^7 \sqrt{\frac{\left(2 r_{h}^4 \left(r_{h} (r_{h}-2 M)+Q^2\right)+\Theta ^2 \left(3 Q^2 r_{h} (10 M-3 r_{h})+M r_{h}^2 (4 r_{h}-11 M)-14 Q^4\right)\right) \left(\zeta  \Theta ^2+2 r_{h}^4 \left(r_{h} (r_{h}-2 M)+Q^2\right)\right)}{r_{h}^8 \left(r_{h} (r_{h}-2 M)+Q^2\right)^2}}},
\end{split}
\fe
where $\zeta$
\ie
\zeta =3 Q^2 r_{h} (r_{h}-2 M)+M r_{h}^2 (3 M-2 r_{h})+2 Q^4.
\fe
and considering small values for $\Theta$ and $Q$, it turns out to be

\begin{align}
\label{approooo}
T_{H}& \approx   \, \, \frac{M}{2 \pi  r_{h}^2} -\frac{Q^2}{2 \left(\pi  r_{h}^3\right)} \\ \nonumber
& +\left[\frac{Q^2 \left(-272 M^3+349 M^2 r_{h}-143 M r_{h}^2+18 r_{h}^3\right)}{4 \pi  r_{h}^6 (r_{h}-2 M)^2}+\frac{-40 M^3+33 M^2 r_{h}-6 M r_{h}^2}{4 \pi  r_{h}^5 (r_{h}-2 M)}\right]\Theta ^2.
\end{align}

It is clear that when $\Theta$ goes to zero, the Hawking temperature recovers the Reissner--Nordstr\"{o}m black hole. In the next step, we show the Hawking temperature as a function of the event horizon. To do so, we consider the solution of the mass coming from $1/g_{rr}^{\scriptscriptstyle {(\Theta},\scriptscriptstyle{Q)}}(r) = 0$, which leads to
\ie
\label{massadds}
r_h=M+\sqrt{M^2-Q^2}\quad\quad \text{or}\quad\quad M = \,  \frac{Q^2+r_{h}^2}{2 r_{h}}.
\fe
It is worth to notice that the horizon radius keeps the same as the commutative model of the Reissner--Nordstr\"{o}m black hole. Therefore, after substituting mass from Eq. \eqref{massadds} in Eq. \eqref{approooo}, we obtain
\ie
\label{ronly}
T_{H} \approx \frac{1}{4 \pi  r_{h}} -\frac{Q^2}{4 \pi  r_{h}^3} -\frac{2 \Theta ^2 Q^2}{\pi  r_{h}^5}-\frac{\Theta ^2}{8 \pi  Q^2 r_{h}}+\frac{5 \Theta ^2}{8 \pi  r_{h}^3}.
\fe

\begin{figure}
    \centering
     \includegraphics[scale=0.6]{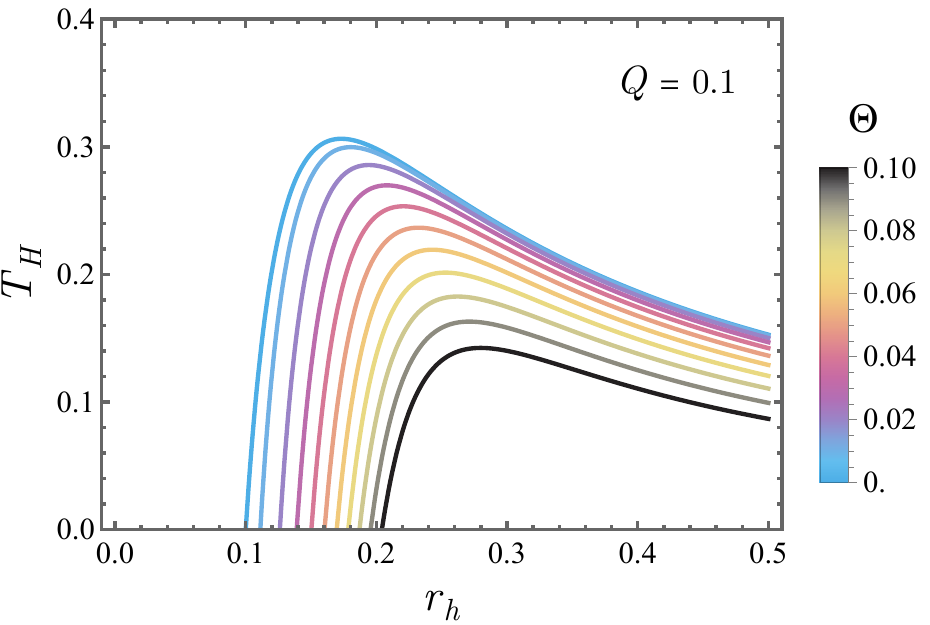}
    \caption{The Hawking temperature is plotted as a function of the event horizon radius $r_{h}$ for $Q=0.1$ and various values of $\Theta$.}
    \label{hawkingtemperature}
\end{figure}

In Fig. \ref{hawkingtemperature}, the behavior of the Hawking temperature concerning the event horizon radius $r_{h}$ is depicted for several values of $\Theta$, assuming fixed values $Q = 0.5$. It is evident that increasing the NC parameter results in a reduction in the Hawking temperature for this specific configuration.

\begin{figure}
    \centering
     \includegraphics[scale=0.6]{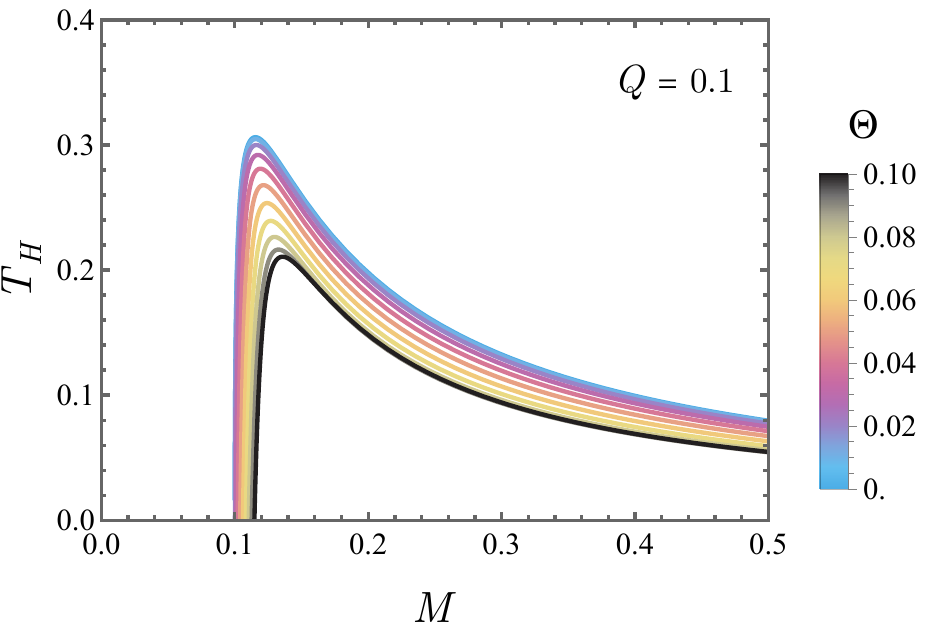}
    \caption{The Hawking temperature is plotted with respect to the mass $M$ for different values of $\Theta$, while the charge is set to $Q=0.1$.}
    \label{hawkingtemperaturemass}
\end{figure}

In addition, it would be important to investigate this thermal quantity concerning the mass $M$. Using the expression of the event horizon present in Eq. \eqref{massadds} and substituting it in the Hawking temperature expression in Eq. \eqref{ronly}, leads to

\begin{align} \nonumber\label{Tmass}
T_{H} \approx &  -\frac{2 \Theta ^2 Q^2}{\pi  \left(\sqrt{M^2-Q^2}+M\right)^5}-\frac{\Theta ^2}{8 \pi  Q^2 \left(\sqrt{M^2-Q^2}+M\right)}+\frac{5 \Theta ^2}{8 \pi  \left(\sqrt{M^2-Q^2}+M\right)^3}\\
& -\frac{Q^2}{4 \pi  \left(\sqrt{M^2-Q^2}+M\right)^3}+\frac{1}{4 \pi  \left(\sqrt{M^2-Q^2}+M\right)}.
\end{align}
Fig. \ref{hawkingtemperaturemass} depicts the Hawking temperature concerning the mass and the existence of remnant masses becomes apparent, as the Hawking temperature tends to zero while the mass remains finite, i.e., $T_{H} \to 0$ as $M \to M_{\text{rem}} \neq 0$. To determine an explicit expression for the remnant mass $M_{\text{rem}}$, we consider Eq. \eqref{Tmass} in the regime of small $\Theta$ and $Q$, which yields the following approximation

\begin{equation}
M_{\text{rem}}  \approx \, Q + \frac{9\Theta^{4}}{2 Q^3}. 
\end{equation}

In other words, the above expression indicates that the black hole does not undergo complete evaporation; instead, it leaves behind a nonzero remnant mass, $M_{\text{rem}} \neq 0$.

\subsection{Heat capacity}

In this section, we conclude our analysis by examining the behavior of the heat capacity by considering its related expression to entropy $S$ and Hawking temperature as follows
\begin{align}\label{eq:cv}
C_{V} &=T_H 
\frac{\partial S}{\partial T_H}\\ \nonumber
&=
\frac{4 \pi  Q^2 {r_h}^6 (Q-{r_h}) (Q+{r_h})+2 \pi  \Theta ^2 {r_h}^2 \left(58 Q^6+16 Q^4 {r_h}^2-5 Q^2 {r_h}^4+{r_h}^6\right)}{-6 Q^4 {r_h}^4+2 Q^2 {r_h}^6-\Theta ^2 \left(406 Q^6+80 Q^4 {r_h}^2-15 Q^2 {r_h}^4+{r_h}^6\right)} 
\end{align}

\begin{figure}
    \centering
     \includegraphics[scale=0.6]{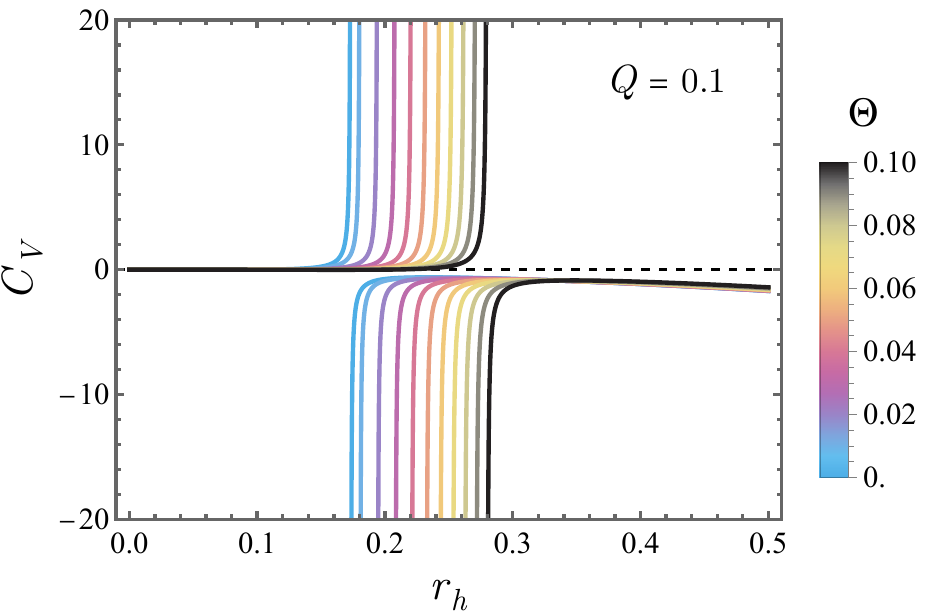}
    \caption{The heat capacity $C_{V}$ is shown as a function of the event horizon $r_{h}$ for different values of the charge $\Theta$, with the parameters set fixed at $Q=0.1$.}
    \label{fig:heatcapp}
\end{figure}

In Fig. \ref{fig:heatcapp}, the heat capacity $C_{V}$ is displayed as a function of the event horizon radius $r_{h}$ for different values of the $\Theta$, with charge fixed at $Q = 0.1$. This plot highlights phase transitions and the regions where $C_{V}$ assumes positive or negative values.

\begin{figure}
    \centering
     \includegraphics[width=75mm]{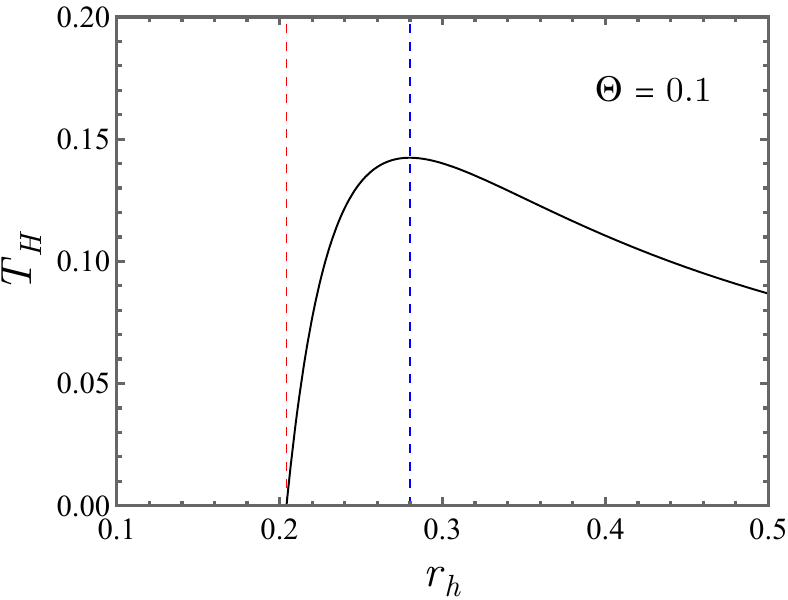}
     \includegraphics[width=75mm]{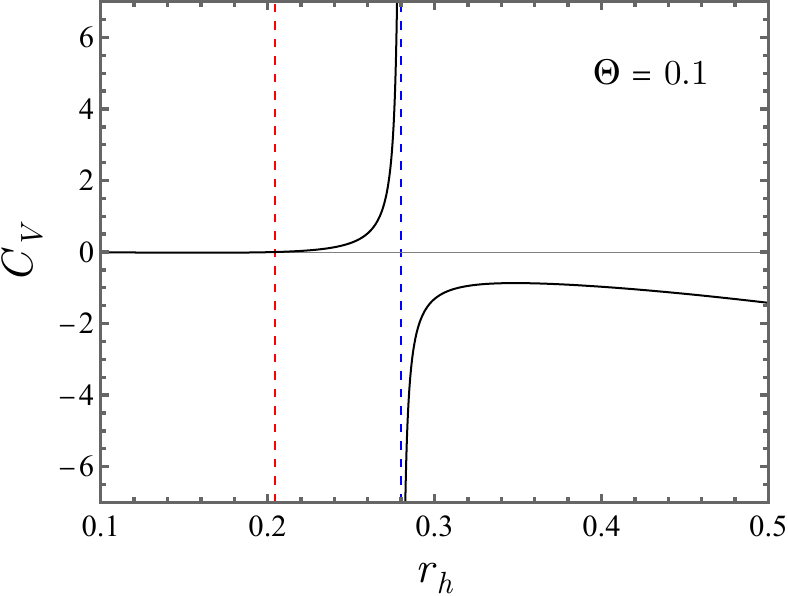}
    \caption{The Hawking temperature and heat capacity are shown as a function of the event horizon $r_{h}$ for fixed values of the $Q = 0.1$ and $\Theta = 0.1$. The red and blue dashed lines correspond to the remnant and phase transition radius.}
    \label{CompTCV}
\end{figure}
 The thermodynamic consistency between the Hawking temperature $T_H$ and the heat capacity at constant volume $C_V$ is a crucial subject in black hole physics. As derived from the fundamental relation in Eq. \ref{eq:cv}, the heat capacity vanishes precisely at the radius where the Hawking temperature goes to zero $(T_H \rightarrow 0)$, indicating a thermodynamically stable remnant. Furthermore, the heat capacity diverges at the extrema of the temperature profile, where $(\frac{\partial T_H}{\partial {r_h}} \rightarrow 0)$, signaling a second-order phase transition. For better visualization, both the Hawking temperature and the corresponding heat capacity are shown together in Fig.~\ref{CompTCV}. The remnant radius and the phase transition point are highlighted with red and blue dashed lines, respectively, providing a direct visual confirmation of the aforementioned thermodynamic relationships.

\section{Hawking radiation}

In this section, we develop the earlier examination, which confirmed the thermal emission of the black hole via the Hawking temperature, to explore the quantum radiation itself. We focus on both bosonic and fermionic particle modes. To derive analytical expressions, we adopt the same angular components as in the Reissner--Nordstr\"{o}m metric. The corresponding particle creation densities are calculated for each case, enabling a comparative analysis of the emission rates. As will be shown, for a given frequency $\omega$, the emission of bosons surpasses that of fermions.


\subsection{Bosons}

To conduct our analysis, we examine the following metric tensor configuration
\ie
\label{eqwerer}
\mathrm{d}s^{2} =  g_{tt}^{\scriptscriptstyle {(\Theta},\scriptscriptstyle{Q)}}(r) \mathrm{d}t^{2} +  g_{rr}^{\scriptscriptstyle {(\Theta},\scriptscriptstyle{Q)}} (r) \mathrm{d}r^{2} + g_{\theta\theta}^{\scriptscriptstyle {(\Theta},\scriptscriptstyle{Q)}}(r,\theta)\mathrm{d}\theta^{2} + g_{\varphi\varphi}^{\scriptscriptstyle {(\Theta},\scriptscriptstyle{Q)}}(r,\theta) \mathrm{d}\varphi^{2}.
\fe
Under the Hamilton--Jacobi approach, the equation characterizing the radial motion of a massless particle takes the form \cite{Filho:2023qxu,Filho:2023voz,vanzo2011tunnelling}
\ie
\frac{1}{g_{tt}^{\scriptscriptstyle {(\Theta},\scriptscriptstyle{Q)}}(r)}(\partial_t I)^2 + \frac{1}{g_{rr}^{\scriptscriptstyle {(\Theta},\scriptscriptstyle{Q)}}(r)}(\partial_r I)^2=0\,.
\label{m2}
\fe

As demonstrated in the following analysis, the classical action admits a representation in which the positive and negative signs correspond to outgoing and ingoing particles, respectively.
\ie
I_{\pm}=-\omega t\pm\int \omega \frac{\mathrm{d}r}{\sqrt{\frac{|g_{tt}^{\scriptscriptstyle {(\Theta},\scriptscriptstyle{Q)}}(r)|}{g_{rr}^{\scriptscriptstyle {(\Theta},\scriptscriptstyle{Q)}}(r)}}}\,,
\label{m3}
\fe
Here, $\omega = - \partial_t I$ corresponds to the Killing energy. Through near-horizon expansion, we obtain
\ie
g_{tt}^{\scriptscriptstyle {(\Theta},\scriptscriptstyle{Q)}}(r) =  \left.
\frac{\mathrm{d}}{\mathrm{d}r}\Bigl(g_{tt}^{\scriptscriptstyle {(\Theta},\scriptscriptstyle{Q)}}(r)\Bigr)
\right|_{r=r_{h}}  (r-r_{ h})+ \dots \;,\quad  \frac{1}{g_{rr}^{\scriptscriptstyle {(\Theta},\scriptscriptstyle{Q)}}(r)} 
= 
\left.
\frac{\mathrm{d}}{\mathrm{d}r}\Bigl(\tfrac{1}{g_{rr}^{\scriptscriptstyle {(\Theta},\scriptscriptstyle{Q)}}(r)}\Bigr)
\right|_{r=r_{h}}
\, (r-r_{h}) 
+ \dots
\;,
\fe
and utilizing Feynman's method, we arrive at
\ie
\mbox{Im}\!\int\!\mathrm{d}I_+-\mbox{ Im}\!\int\! \mathrm{d}I_-=\frac{\pi\omega}{\kappa},
\fe
where
\ie\label{kvw}
\kappa=\frac{1}{2}\sqrt{\left.
\frac{\mathrm{d}}{\mathrm{d}r}\Bigl(g_{tt}^{\scriptscriptstyle {(\Theta},\scriptscriptstyle{Q)}}(r)\Bigr)
\right|_{r=r_{h}}   \left.
\frac{\mathrm{d}}{\mathrm{d}r}\Bigl(\tfrac{1}{g_{rr}^{\scriptscriptstyle {(\Theta},\scriptscriptstyle{Q)}}(r)}\Bigr)
\right|_{r=r_{h}}},
\fe
is, therefore, the so--called surface gravity. If we consider $\Theta$ small, we obtain
\ie
\kappa \approx \, \, \frac{M r_{h}-Q^2}{r_{h}^3} + \frac{\Theta ^2 \left(28 M^2 r_{h}^2-90 M Q^2 r_{h}-9 M r_{h}^3+48 Q^4+24 Q^2 r_{h}^2\right)}{4 r_{h}^7}.
\fe

By defining 
\ie
\Gamma = e^{-\frac{2 \pi  \omega }{k}},
\fe
From this analysis, we derive the bosonic particle creation density $n_b^{(\Theta,Q)}$ explicitly as
\ie
\begin{split}
n_b^{(\Theta,Q)} & = \frac{\Gamma}{1-\Gamma} = \frac{1}{e^{\frac{8 \pi  \omega  \left(\sqrt{M^2-Q^2}+M\right)^7}{\Lambda }}-1},
\end{split}
\fe
where
\ie
\begin{split}
\Lambda \equiv & \, \, 64 M^6+4 M^4 \left(5 \Theta ^2-28 Q^2\right)+M Q^2 \sqrt{M^2-Q^2} \left(20 Q^2-33 \Theta ^2\right)+M^2 \left(52 Q^4-43 \Theta ^2 Q^2\right) \\
& +64 M^5 \sqrt{M^2-Q^2}+20 M^3 \sqrt{M^2-Q^2} \left(\Theta ^2-4 Q^2\right)-4 Q^4 \left(Q^2-6 \Theta ^2\right).
\end{split}
\fe

\begin{figure}
    \centering
    \includegraphics[scale=0.6]{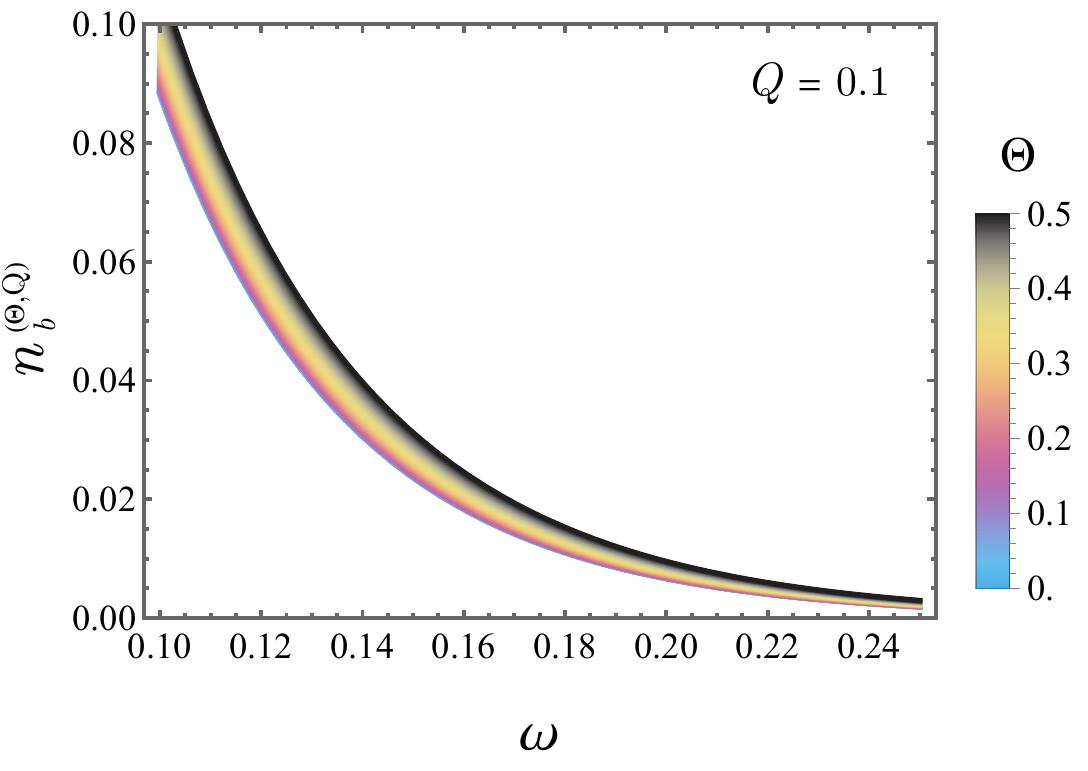}
    \caption{The particle creation density $n^{\scriptscriptstyle {(\Theta},\scriptscriptstyle{Q)}}$ is presented as a function of the frequency $\omega$ for a fixed value of $Q = 0.1$ and $M=1$ and different values of $\Theta$.}
    \label{partbon}
\end{figure}

Fig. \ref{partbon} displays the particle creation density $n^{\scriptscriptstyle {(\Theta},\scriptscriptstyle{Q)}}$ as a function of the frequency $\omega$, with fixed parameters $Q = 0.1$ and $M = 1$, while varying the values of $\Theta$. Overall, the magnitude of $n^{\scriptscriptstyle {(\Theta},\scriptscriptstyle{Q)}}$ increases with larger values of $\Theta$.

\subsection{Fermions}

It is important to emphasize that the subsequent calculations are performed under the assumption that backreaction effects are neglected. Within the context of quantum tunneling, particle emission from a black hole is modeled as a quantum process wherein particles traverse the event horizon. The likelihood of such tunneling events can be determined using the methodologies outlined in Refs. \cite{angheben2005hawking,kerner2006tunnelling,kerner2008fermions}, along with the associated literature referenced therein.


Black hole radiation arises from their inherent temperature, in a manner analogous to blackbody radiation. Nevertheless, the emitted spectrum is subject to modification by greybody factors, which alter the characteristics of the outgoing radiation. The spectrum is anticipated to comprise particles of various spin types, including fermions. Foundational work in Ref. \cite{o69}, along with subsequent investigations \cite{o75,o72,o71,o74,o73,o70}, has demonstrated that massless bosons and fermions radiate at an identical temperature. Furthermore, studies of spin--1 bosons indicate that the Hawking temperature remains unaffected, even when quantum corrections beyond the semiclassical approximation—specifically those of higher--order in $\hbar$ are taken into account \cite{o77,o76}.

The behaviour of fermions is frequently characterized by the phase of the spinor wave function, which satisfies the Hamilton--Jacobi equation. An alternative formulation of the action, as proposed in \cite{o84,o83,vanzo2011tunnelling}, is expressed as
\ie
\mathcal{S}^{(\Psi)} = S^{(0)} + \psi^{(\uparrow \downarrow)},
\fe
where $S_0$ denotes the classical action for scalar particles and $\psi^{(\uparrow \downarrow)}$ accounts for the spin corrections. These spin--dependent terms arise from the coupling between the fermion’s intrinsic spin and the background spin connection, and they ensure the regularity of the solution at the event horizon. Given that such corrections primarily influence spin precession and are typically small in magnitude, they are neglected in the present analysis. Additionally, the contribution of emitted particles’ spin to the total angular momentum of the black hole is negligible--particularly for non--rotating black holes with masses significantly greater than the Planck mass \cite{vanzo2011tunnelling}. On average, emissions of particles with opposite spin orientations occur symmetrically, resulting in no net change to the black hole’s angular momentum.

This study examines the tunneling process of fermionic particles as they traverse the event horizon of a specific black hole spacetime. Alternative approaches, including those based on generalized Painlev\'{e}--Gullstrand and Kruskal--Szekeres coordinate systems, are explored in the seminal work \cite{o69}. The analysis commences with a general form of the spacetime metric, as given in Eq.~\eqref{eqwerer}. In a curved spacetime background, the behavior of fermions is governed by the Dirac equation, which takes the form
\ie
\left(\Tilde{\gamma}^\mu \nabla_\mu + m \right) \Psi(x) = 0,
\fe
with
\ie
\nabla_\mu = \partial_\mu + \frac{\mathbbm{i}}{2} {\Gamma^\alpha_{\;\mu}}^{\;\beta} \,\Tilde{\Sigma}_{\alpha\beta},\fe
and 
\ie
\Tilde{\Sigma}_{\alpha\beta} = \frac{\mathbbm{i}}{4} [\Tilde{\gamma}_\alpha,  \Tilde{\gamma}_\beta].
\fe

It is worth emphasizing that the coordinates are represented as \( x^\mu \equiv (t, r, \theta, \varphi) \). The matrices \( \Tilde{\gamma}^\mu \) satisfy the defining relations of the Clifford algebra, given by
\begin{equation}
\{ \Tilde{\gamma}^\mu, \Tilde{\gamma}^\nu \} = 2 g^{\mu\nu} \mathbb{I},
\end{equation}

with $\mathbbm{1}$ represents the $4 \times 4$ identity matrix. Based upon this formulation, the $\Tilde{\gamma}$ matrices are prescribed as given below

\begin{eqnarray*}
 \Tilde{\gamma}^{t} &=&\frac{\mathbbm{i}}{\sqrt{|g_{tt}^{\scriptscriptstyle {(\Theta},\scriptscriptstyle{Q)}}|}}\left( \begin{array}{cc}
\bf{1}& \bf{ 0} \\ 
\bf{ 0} & -\bf{ 1}%
\end{array}%
\right), \;\;
\Tilde{\gamma}^{r} =\sqrt{\frac{1}{g_{rr}^{\scriptscriptstyle {(\Theta},\scriptscriptstyle{Q)}}}}\left( 
\begin{array}{cc}
\bf{0} &  \Tilde{\sigma}_{3} \\ 
 \Tilde{\sigma}_{3} & \bf{0}%
\end{array}%
\right), \\
\Tilde{\gamma}^{\theta } &=&\frac{1}{\sqrt{g_{\theta\theta}^{\scriptscriptstyle{(\Theta},\scriptscriptstyle{Q)}}}}\left( 
\begin{array}{cc}
\bf{0} &  \Tilde{\sigma}_{1} \\ 
 \Tilde{\sigma}_{1} & \bf{0}%
\end{array}%
\right), \;\;
\Tilde{\gamma}^{\varphi } =\frac{1}{\sqrt{g_{\varphi\varphi}^{\scriptscriptstyle{(\Theta},\scriptscriptstyle{Q)}}}}\left( 
\begin{array}{cc}
\bf{0} &  \Tilde{\sigma}_{2} \\ 
 \Tilde{\sigma}_{2} & \bf{0}%
\end{array}%
\right).
\end{eqnarray*}%
In this context, $\Tilde{\sigma}$ corresponds to the Pauli matrices, which obey the conventional commutation relation  $\Tilde{\sigma}_i \Tilde{\sigma}_j = \mathbf{1} \delta_{ij} + \mathbbm{i} \varepsilon_{ijk} \Tilde{\sigma}_k, \quad \text{where} \quad i,j,k =1,2,3$. 
Additionally, the matrix associated with $\Tilde{\gamma}^5$ can be equivalent to
\begin{equation*}
\Tilde{\gamma}^{5} = \mathbbm{i} \Tilde{\gamma}^{t}\Tilde{\gamma}^{r}\Tilde{\gamma}^{\theta }\Tilde{\gamma}^{\varphi } = \frac{\mathbbm{i}}{\sqrt{{g_{tt}^{\scriptscriptstyle {(\Theta},\scriptscriptstyle{Q)}} \, g_{rr}^{\scriptscriptstyle {(\Theta},\scriptscriptstyle{Q)}}g_{\theta\theta}^{\scriptscriptstyle{(\Theta},\scriptscriptstyle{Q)}}{g_{\varphi\varphi}^{\scriptscriptstyle{(\Theta},\scriptscriptstyle{Q)}}}}}}\left( 
\begin{array}{cc}
\bf{ 0} & - \bf{ 1} \\ 
\bf{ 1} & \bf{ 0}%
\end{array}%
\right)\:.
\end{equation*}

To describe a Dirac field with its spin aligned upward along the positive $r$--axis, the adopted ansatz is given by \cite{vagnozzi2022horizon}
\begin{equation}
\Psi^{(+)}(x) = \left( \begin{array}{c}
\Tilde{H}(x) \\ 
0 \\ 
\Tilde{Y}(x) \\ 
0
\end{array}
\right) \exp \left[ \mathbbm{i} \, \psi^{(+)}(x)\right]\;.
\label{spinupbh} 
\end{equation}

This study focuses on the spin--up $(+)$ configuration, while the spin--down $(-)$ case, oriented along the negative $r$--axis, follows a similar treatment. Substituting the ansatz \eqref{spinupbh} into the Dirac equation and following Vanzo et al. \cite{vanzo2011tunnelling}, we keep only the leading--order terms in $\hbar$
\begin{align}
&-\frac{\mathbbm{i}}{\sqrt{|g_{tt}^{\scriptscriptstyle {(\Theta},\scriptscriptstyle{Q)}}(r)|}}\left( \Tilde{H}(x)\,\partial_{t}\psi^{(+)}\right)
-\sqrt{\frac{1}{g_{rr}^{\scriptscriptstyle {(\Theta},\scriptscriptstyle{Q)}}(r)}}\left( \Tilde{Y}(x)\,\partial_{r}\psi^{(+)}\right)
+m \mathbbm{i} \Tilde{H}(x) = 0, \label{eq1} \\[10pt]
& - \frac{1}{\sqrt{g_{\theta\theta}^{\scriptscriptstyle {(\Theta},\scriptscriptstyle{Q)}}(r,\theta)}}\left(\Tilde{Y}(x)\,\partial_{\theta}\psi^{(+)}\right)
-\frac{1}{\sqrt{g_{\varphi\varphi}^{\scriptscriptstyle {(\Theta},\scriptscriptstyle{Q)}}(r,\theta)}}\left( \mathbbm{i}\Tilde{Y}(x)\,\partial_{\phi}\psi^{(+)}\right) = 0, \label{eq2} \\[10pt]
&    \frac{\mathbbm{i}}{\sqrt{|g_{tt}^{\scriptscriptstyle {(\Theta},\scriptscriptstyle{Q)}}(r)|}}\left(\Tilde{Y}(x)\,\partial_{t}\psi^{(+)}\right)
-\sqrt{\frac{1}{g_{rr}^{\scriptscriptstyle {(\Theta},\scriptscriptstyle{Q)}}(r)}}\left(\Tilde{H}(x)\,\partial_{r}\psi^{(+)}\right)
+ m \mathbbm{i} \Tilde{Y}(x) = 0, \label{eq3} \\[10pt]
& - \frac{1}{\sqrt{g_{\theta\theta}^{\scriptscriptstyle {(\Theta},\scriptscriptstyle{Q)}}(r,\theta)}}\left( \Tilde{H}(x)\,\partial_{\theta}\psi^{(+)}\right)
-\frac{\mathbbm{i}}{\sqrt{g_{\varphi\varphi}^{\scriptscriptstyle {(\Theta},\scriptscriptstyle{Q)}}(r,\theta)}}\left( \Tilde{H}(x)\,\partial_{\phi}\psi^{(+)}\right) = 0. \label{eq4}
\end{align}

and if the action takes the form of
\ie
\psi^{(+)}= - \omega\, t + \Tilde{\chi}(r) + L(\theta ,\varphi ) 
\fe
The following equations will be derived
\cite{vanzo2011tunnelling} 

\begin{align}
&+\frac{\mathbbm{i}\, \omega \Tilde{H}(x)}{\sqrt{|g_{tt}^{\scriptscriptstyle {(\Theta},\scriptscriptstyle{Q)}}(r)|}} 
-\sqrt{\frac{1}{g_{rr}^{\scriptscriptstyle {(\Theta},\scriptscriptstyle{Q)}}(r)}} \Tilde{Y}(x)\,\Tilde{\chi}^{\prime}(r)
+m \mathbbm{i} \Tilde{H}(x) = 0, \label{eq11} \\[10pt]
& - \Tilde{Y}(x) \left(  \frac{\partial_{\theta}L(\theta,\varphi)}{\sqrt{g_{\theta\theta}^{\scriptscriptstyle {(\Theta},\scriptscriptstyle{Q)}}(r,\theta)}}
+\frac{\mathbbm{i}\,\partial_{\phi} L(\theta,\varphi)}{\sqrt{g_{\varphi\varphi}^{\scriptscriptstyle {(\Theta},\scriptscriptstyle{Q)}}(r,\theta)}} \right) = 0, \label{eq21} \\[10pt]
&    -\frac{\mathbbm{i} \, \omega \Tilde{Y}(x)}{\sqrt{|g_{tt}^{\scriptscriptstyle {(\Theta},\scriptscriptstyle{Q)}}(r)|}}
-\sqrt{\frac{1}{g_{rr}^{\scriptscriptstyle {(\Theta},\scriptscriptstyle{Q)}}(r)}}\Tilde{H}(x)\Tilde{\chi}^{\prime}(r)
+ m \mathbbm{i} \Tilde{Y}(x) = 0, \label{eq31} \\[10pt]
& - \Tilde{H}(x) \left(  \frac{\partial_{\theta}L(\theta,\varphi)}{\sqrt{g_{\theta\theta}^{\scriptscriptstyle {(\Theta},\scriptscriptstyle{Q)}}(r,\theta)}}
+\frac{\mathbbm{i}\,\partial_{\phi} L(\theta,\varphi)}{\sqrt{g_{\varphi\varphi}^{\scriptscriptstyle {(\Theta},\scriptscriptstyle{Q)}}(r,\theta)}} \right) = 0. \label{eq41}
\end{align}

The explicit forms of \( \Tilde{H}(x) \) and \( \Tilde{Y}(x) \) do not influence the conclusion that Eqs.~\eqref{eq21} and \eqref{eq41} yield the constraint \( L_{\theta} + \mathbbm{i}(\sin \theta)^{-1} L_{\varphi} = 0 \), indicating that \( L(\theta, \varphi) \) must be a complex-valued function. This condition applies to both outgoing and ingoing modes. As a result, in calculating the ratio of outgoing to ingoing tunneling probabilities, the contributions from \( L \) cancel out, allowing it to be excluded from further consideration \cite{vanzo2011tunnelling}. For massless particles, Eqs.~\eqref{eq11} and \eqref{eq31} admit two linearly independent solutions

\ie
\Tilde{H}(x) = - \mathbbm{i} \Tilde{Y}, \qquad \Tilde{\chi}^{\prime }(r) = \Tilde{\chi}_{\text{out}}' = \frac{\omega}{\sqrt{\frac{g_{tt}^{\scriptscriptstyle {(\Theta},\scriptscriptstyle{Q)}}(r)}{g_{rr}^{\Theta,l}(r)}}},
\fe
\ie
\Tilde{H}(x) = \mathbbm{i} \Tilde{Y}(x), \qquad \Tilde{\chi}^{\prime }(r) = \Tilde{\chi}_{\text{in}}' = - \frac{\omega}{\sqrt{\frac{g_{tt}^{\scriptscriptstyle {(\Theta},\scriptscriptstyle{Q)}}(r)}{g_{rr}^{\Theta,l}(r)}}}.
\fe
In this framework, \( \Tilde{\chi}_{\text{out}} \) and \( \Tilde{\chi}_{\text{in}} \) correspond to the solutions describing outgoing and incoming particles, respectively~\cite{vanzo2011tunnelling}. Accordingly, the total tunneling probability is given by \( \Tilde{\Gamma}_{\psi} \sim e^{-2\, \text{Im} \, (\Tilde{\chi}_{\text{out}} - \Tilde{\chi}_{\text{in}})} \). Thus,

\ie
 \Tilde{\chi}_{ \text{out}}(r)= -  \Tilde{\chi}_{ \text{in}} (r) = \int \mathrm{d} r \,\frac{\omega}{\sqrt{\frac{g_{tt}^{\scriptscriptstyle {(\Theta},\scriptscriptstyle{Q)}}(r)}{g_{rr}^{\Theta,l}(r)}}}\:.
\fe

A key aspect is that the dominant energy condition, together with Einstein’s field equations, guarantees that $g_{tt}^{\scriptscriptstyle {(\Theta},\scriptscriptstyle{Q)}}(r)$ and $1/g_{rr}^{\scriptscriptstyle {(\Theta},\scriptscriptstyle{Q)}}(r)$ share the same zeros. Near $r = r_{h}$, these metric components exhibit a linear behavior, revealing the presence of a simple pole with a well--defined coefficient. By utilizing Feynman’s method, the following expression is found
\ie
2\mbox{Im}\;\left(  \Tilde{\chi}_{ \text{out}} -  \Tilde{\chi}_{ \text{in}} \right) =\mbox{Im}\int \mathrm{d} r \,\frac{4\omega}{\sqrt{\frac{g_{tt}^{\scriptscriptstyle {(\Theta},\scriptscriptstyle{Q)}}(r)}{g_{rr}^{\Theta,l}(r)}}}=\frac{2\pi\omega}{\kappa}.
\fe
Within this framework, the particle number density \( n_{f}^{(\Theta,Q)} \) associated with the specified black hole solution is governed by the relation
 $\Tilde{\Gamma}_{\psi} \sim e^{-\frac{2 \pi \omega}{\kappa}}$
\ie
n_{f}^{(\Theta,Q)} = \frac{1}{e^{\frac{2 \pi  \omega }{\kappa}}+1}.
\fe

\begin{figure}
    \centering
    \includegraphics[scale=0.6]{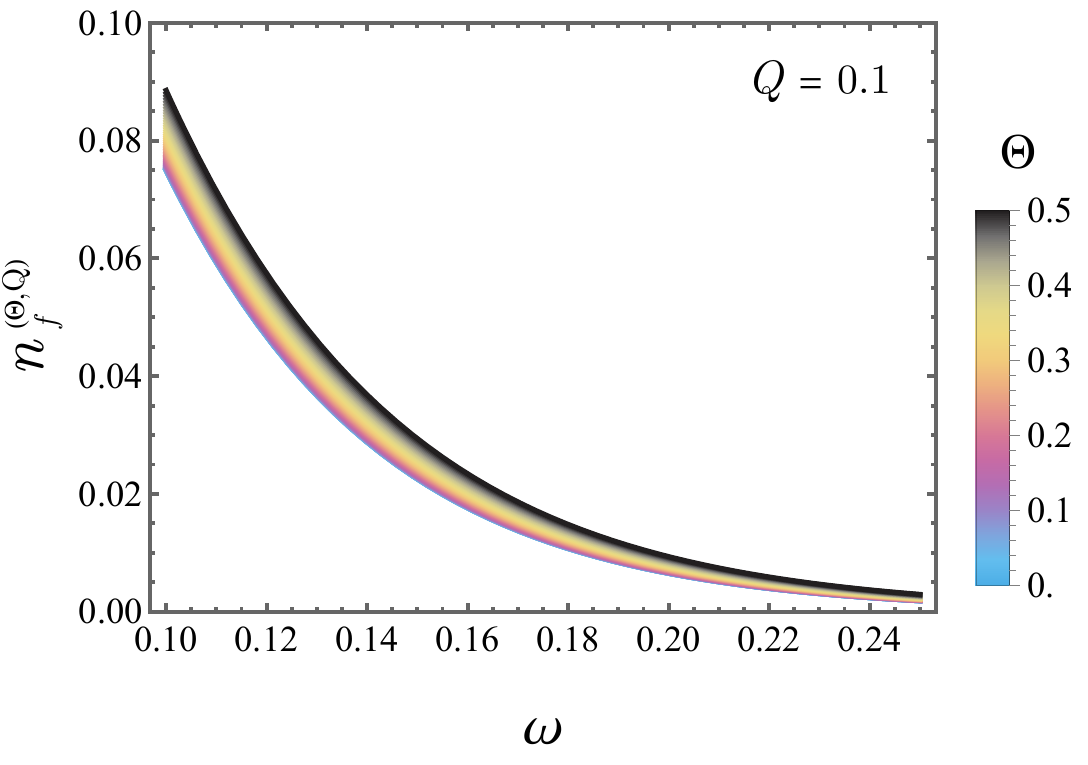}
    \caption{The particle creation density $n_{\psi}$ is presented as a function of the frequency $\omega$ for different values of the NC parameter $\Theta$, with fixed values of $Q = 0.1$ and $M = 1$.}
    \label{partferm}
\end{figure}

Figure~\ref{partferm} presents the particle creation density for fermions, \( n^{(\Theta ,Q)} \). Similar to the bosonic case, the NC parameter \( \Theta \) leads to an enhancement in the magnitude of the particle production density.

\begin{figure}
    \centering
    \includegraphics[scale=0.6]{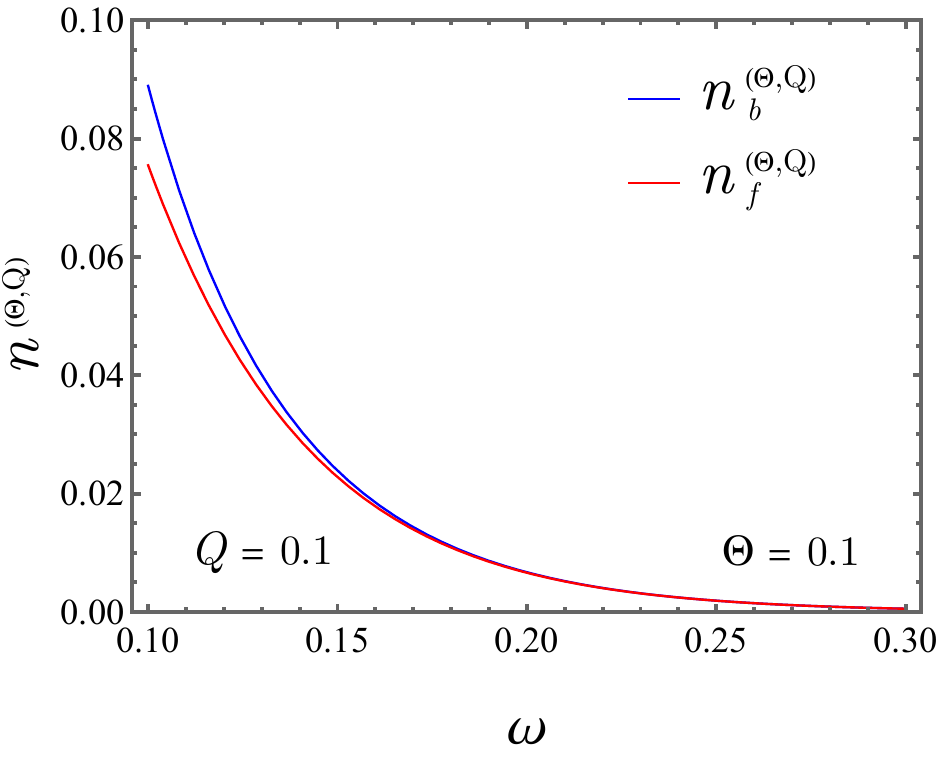}
    \caption{The particle creation comparison (for bosons and fermions) is illustrated with respect to the frequency $\omega$, while keeping $Q = 0.1$, $M = 1$, and $\Theta = 0.1$ fixed.}
    \label{compartt}
\end{figure}

Additionally, Fig. \ref{compartt} presents a comparison of the particle creation densities for bosons and fermions. Overall, the bosonic case exhibits a greater magnitude at lower frequencies compared to the fermionic case.

\FloatBarrier
\section{Scalar perturbation}

This section examines the massless
scalar perturbation through the non--commutative Reissner--Nordstr\"{o}m black hole spacetime. 
\begin{equation}\label{klein}
\Box\psi=\frac{1}{{\sqrt { - g} }}{\partial _\mu }(\sqrt { - g} {g^{\mu \nu }}{\partial _\nu }\psi ) = 0 .
\end{equation}

Due to the complex form of the metric, the Klein--Gordon equation can not be solved with a common numerical method. Here we follow the method introduced by Ref.~\cite{chen2022eikonal}. This novel method proposes considering the deformed stationary and axisymmetric metric in a perturbation form in which the deformation is controlled by a small dimensionless modification parameter \( \epsilon \). According to this approach, the modified metric can be expressed via a correction term added to the main metric as $ g_{ij}^{{\scriptscriptstyle {\text{mod}}}} = g_{ij} + \epsilon h_{ij}$. Here, $g_{ij}^{{\scriptscriptstyle {\text{mod}}}}$ and $g_{ij}$ represent the metric incorporated into the modified and standard metric, respectively. Also, $h_{ij}$ denotes the correction coefficients. Building upon the approach introduced in Ref. \cite{zhao2023quasinormal}, a key assumption in this part is that the main metric function is Reissner--Nordstr\"{o}m and the deformation caused by noncommutativity. Now the metric function in Eq. \eqref{metric1} can be rewritten as 
\begin{subequations}\label{modmetrics}
    \begin{align}
{{ g^{\scriptscriptstyle {(\Theta},\scriptscriptstyle{Q)}}_{00}}} &= - f(r)(1 + \epsilon{A_j}{{\mathop{\rm cos}\nolimits} ^j}\theta ),\\
{g^{\scriptscriptstyle {(\Theta},\scriptscriptstyle{Q)}}_{11}} &= {f(r)^{ - 1}}(1 + \epsilon{B_j}{{\mathop{\rm cos}\nolimits} ^j}\theta ),\\
{g^{\scriptscriptstyle {(\Theta},\scriptscriptstyle{Q)}}_{22}}& = {r^2}(1 + \epsilon{C_j}{{\mathop{\rm cos}\nolimits} ^j}\theta ),\\
{g^{\scriptscriptstyle {(\Theta},\scriptscriptstyle{Q)}}_{33}}&= {r^2}{{\mathop{\rm sin}\nolimits} ^2}\theta (1 + \epsilon{D_j}{{\mathop{\rm cos}\nolimits} ^j}\theta ),\\
{g^{\scriptscriptstyle {(\Theta},\scriptscriptstyle{Q)}}_{01}}& = \epsilon{a_j}(r){{\mathop{\rm cos}\nolimits} ^j}\theta ,\quad
{g^{\scriptscriptstyle {(\Theta},\scriptscriptstyle{Q)}}_{12}} = \epsilon{c_j}(r){{\mathop{\rm cos}\nolimits} ^j}\theta ,\quad\
{g^{\scriptscriptstyle {(\Theta},\scriptscriptstyle{Q)}}_{23}} = \epsilon{e_j}(r){{\mathop{\rm cos}\nolimits} ^j}\theta,\\
{g^{\scriptscriptstyle {(\Theta},\scriptscriptstyle{Q)}}_{02}}& = \epsilon{b_j}(r){{\mathop{\rm cos}\nolimits} ^j}\theta ,\quad
{g^{\scriptscriptstyle {(\Theta},\scriptscriptstyle{Q)}}_{13}} = \epsilon{d_j}(r){{\mathop{\rm cos}\nolimits} ^j}\theta. 
    \end{align}
\end{subequations}

where $f(r)=1-\frac{2M}{r}+\frac{Q^2}{r^2}$ and the deformation is governed by small parameter $\epsilon$ which is NC parameter $\Theta^2$ in our case. 
The corresponding coefficients for the modified metric components are derived as follows

\begin{align}\label{A}
&{A_0} =\frac{3 Q^2 r (10 M-3 r)+M r^2 (4 r-11 M)-14 Q^4}{2 r^4 \left(r (r-2 M)+Q^2\right)},
\\
&{B_0} = \frac{3 Q^2 r (r-2 M)+M r^2 (3 M-2 r)+2 Q^4}{2 r^4 \left(r (r-2 M)+Q^2\right)},\\
&{C_0} =\frac{r^2 \left(64 M^2-32 M r+r^2\right)+18 Q^2 r (3 r-8 M)+57 Q^4}{16 r^4 \left(r (r-2 M)+Q^2\right)},\\
&{D_0} = \frac{r^2 \left(2 M^2-4 M r+r^2\right)+Q^2 r (5 r-8 M)+4 Q^4}{4 r^4 \left(r (r-2 M)+Q^2\right)},\\
&{A_j}={B_j}={C_j}=0 \quad \text{and} \quad
{D_j}= \frac{5({1 + {{\left( { - 1} \right)}^j}})}{{32{r^2}}} \quad \text{for}\quad j>0,\\ \label{abcd}
&{a_j}(r)= {b_j}(r) = {c_j}(r) = {d_j}(r) = 0.
\end{align}

Now, the Klein--Gordon equation can be explored in a new developed perturbation approach. First, the wave function can be decomposed considering two Killing vectors $\partial_t$ and $\partial_{\phi}$ as 

\begin{equation}
\psi  = \int_{ - \infty }^\infty  {d\omega \sum\limits_{m =  - \infty }^\infty  {{e^{im\varphi }}D_{m,\omega }^2{\psi _{m,\omega }}(r,\theta ){e^{ -i\omega t}}} },
\end{equation}

where $D_{m,\omega}^2 \psi_{m,\omega}(r,\theta) = 0$, with $m$ denoting the azimuthal quantum number and $\omega$ representing the frequency of the mode.
The perturbative method can be applied by decomposing the operator $D_{m,\omega }^2{\psi _{m,\omega }}$ up to the first order of $\Theta^2$ \cite{chen2022eikonal,	zhao2023quasinormal,heidari2024exploring}

\begin{equation}
D_{m,\omega }^2 = D_{(0)m,\omega }^2 + {\Theta ^2}D_{(1)m,\omega }^2 .
\end{equation}

Utilizing the metric described in Eq. \eqref{modmetrics}, the following expressions for $D_{(0)m,\omega }$ and $D_{(1)m,\omega }$ are obtained

\begin{align}
D_{(0)m,\omega }^2 &=  - ({\omega ^2} - \frac{{{m^2}f(r)}}{{{r^2}{{{\mathop{\rm sin}\nolimits} }^2}\theta }}) - \frac{f(r)}{{{r^2}}}{\partial _r}({r^2}f(r){\partial _r}) - {{\mathop{\rm cos}\nolimits} ^j}\theta ({\partial _r}({r^2}f(r){\partial _r}))\\
&- \frac{f(r)}{{{r^2}{{{\mathop{\rm sin}\nolimits} }^2}\theta }}{\partial _\theta }({\mathop{\rm sin}\nolimits} \theta {\partial _\theta }),\\ \nonumber
D_{(1)m,\omega }^2 &= \frac{{{m^2}f(r)}}{{{r^2}{{{\mathop{\rm sin}\nolimits} }^2}\theta }}({A_j} - {D_j}){{\mathop{\rm cos}\nolimits} ^j}\theta  - \frac{f(r)}{{{r^2}}}({A_j} - {B_j}){{\mathop{\rm cos}\nolimits} ^j}\theta ({\partial _r}({r^2}f{\partial _r}))\\
&- \frac{{{f^2(r)}}}{{{r^2}}}({{A'}_j} - {{B'}_j} + {{C'}_j} + {{D'}_j}){{\mathop{\rm cos}\nolimits} ^j}\theta {\partial _r} - \frac{f(r)}{{{r^2}}}({A_j} - {C_j}){{\mathop{\rm cos}\nolimits} ^j}\theta (cot\theta {\partial _\theta } + \partial _\theta ^2)\nonumber \\
&- \frac{f(r)}{{2{r^2}}}({A_j} + {B_j} - {C_j} + {D_j}){\partial _\theta }{{\mathop{\rm cos}\nolimits} ^j}\theta {\partial _\theta } - \frac{{2i\omega f(r)}}{r}{a_j}{{\mathop{\rm cos}\nolimits} ^j}\theta (r{\partial _r} + 1).\nonumber 
\end{align}

The function $\psi_{m,\omega}$ admits an expansion in terms of associated Legendre functions $P_{l}^{m}(\cos\theta)$ and corresponding radial functions $\mathcal{R}_{l,m}(r)$, expressed as ${\psi_{m,\omega}} = \sum_{l' = |m|}^{\infty} P_{l'}^{m}(\cos\theta), \mathcal{R}_{l',m}(r)$. Using this form of $\psi_{m,\omega}$ leads to a the Schr\"{o}dinger like differential equation for $\mathcal{R}_{l,m}(r)$ as 

\begin{equation}\label{sailm}
\frac{\partial^2 {\mathcal{R} _{lm}}}{ \partial{{r^*}}^2} + ({\omega ^2}-{V_{\text{eff}}}){\mathcal{R} _{lm}}=0,
\end{equation}

where $r^*$, called the tortoise coordinate, is described as

\begin{equation}
\frac{{\mathrm{d}r}}{{\mathrm{d}{r^*}}} =f(r)\left[1 + \frac{1}{2}{\Theta ^2}b_{lm}^j({A_j} - {B_j})\right],
\end{equation}

and the effective potential is described in a perturbative form as 
	\begin{equation}\label{Veff}
	{V_{\text{eff}}} = {V_{\text{RN}}} +\Theta^2 {V_{\text{NC}}}.
	\end{equation}

In this context, $V_{\text{RN}}$ represents the effective potential associated with the standard Reissner–Nordstr\"{o}m black hole, while $V_{\text{NC}}$ accounts for the NC correction to the potential. We will investigate more about the effective potential in the following section.

\subsection{Effective potential}

Conducting a series of algebraic steps, the effective potential expression is explicitly formulated as
	\ie \label{Veff}
    \begin{split}
		&{V_{\rm {RN}}} = f(r)\left( {\frac{{l\left( {l + 1} \right)}}{{{r^2}}} +\frac{1}{r} \frac{{\mathrm{d}f(r)}}{{\mathrm{d}r}}} \right),\\
		& V_{\rm {NC}}= \frac{f(r)}{r}\frac{{\mathrm{d}f(r)}}{{\mathrm{d}r}}b_{lm}^0\left( {{A_0} - {B_0}} \right) + \Big[\frac{f(r)}{{{r^2}}}\big[a_{lm}^0\left( {{A_0} - {D_0}} \right) - c_{lm}^0\left( {{A_0} - {C_0}}\right) \\ 
		& -\frac{{d_{lm}^0}}{2}({A_0} + {B_0} - {C_0} + {D_0})
		+ \frac{1}{{4{r^2}}}\frac{\mathrm{d}}{{\mathrm{d}{r^*}}}\left(b_{lm}^0{r^2}\frac{\mathrm{d}}{{\mathrm{d}{r^*}}}({A_0} - {B_0} + {C_0} + {D_0})\right) \\ 
		&- \frac{{b_{lm}^0}}{4}\frac{{{\mathrm{d}^2}}}{{\mathrm{d}{r^*}^2}}({A_0} - {B_0})\big] \Big] - \frac{f(r)}{{{r^2}}}\sum\limits_{j = 1}^\infty  {\left(a_{lm}^j + \frac{1}{2}d_{lm}^j \right){D_j}}  +\sum\limits_{j = 1}^\infty  {\frac{1}{{4{r^2}}}\frac{\mathrm{d}}{{\mathrm{d}{r^*}}}\left(b_{lm}^j{r^2}\frac{\mathrm{d}}{{\mathrm{d}{r^*}}}\right){D_j}}, 
	\end{split}
    \fe
where the coefficients $a^j_{lm}$--$d^j_{lm}$ are described in Appendix I. It is worth noticing the important footprint of NC spacetime, that the standard form of the effective potential is solely dependent on the multipole number \( l \), and exhibits no explicit dependence on the azimuthal quantum number \( m \). However, the presence of NC breaks the degeneracy, and the effective potential is dependent on both $l$ and $m$. Moreover, the coefficients of the effective potential $V_{\text{eff}}$ remain invariant under the sign change of the azimuthal number. For instance the effective potential for the case of $l = 1$ and $m = \pm 1$, will read the following expression

\begin{align}
&V_{\rm{eff}}=\frac{2 \left(r (r-2 M)+Q^2\right) \left(r (M+r)-Q^2\right)}{r^6}+\frac{\Theta ^2}{64 r^{10} \left(r (r-2 M)+Q^2\right)}\\ \nonumber
& \big[2 Q^4 r^2 \left(10343 M^2-7254 M r+1208 r^2\right)+Q^2 r^3 \left(-16264 M^3+15784 M^2 r-4696 M r^2+409 r^3\right)\\ \nonumber
&+2 r^4 (2 M-r) \left(902 M^3-491 M^2 r+42 M r^2+r^3\right)+Q^6 r (3773 r-10292 M)+1784 Q^8\big]
\end{align}

\begin{figure}[ht!]
	\centering
	\includegraphics[width=52mm]{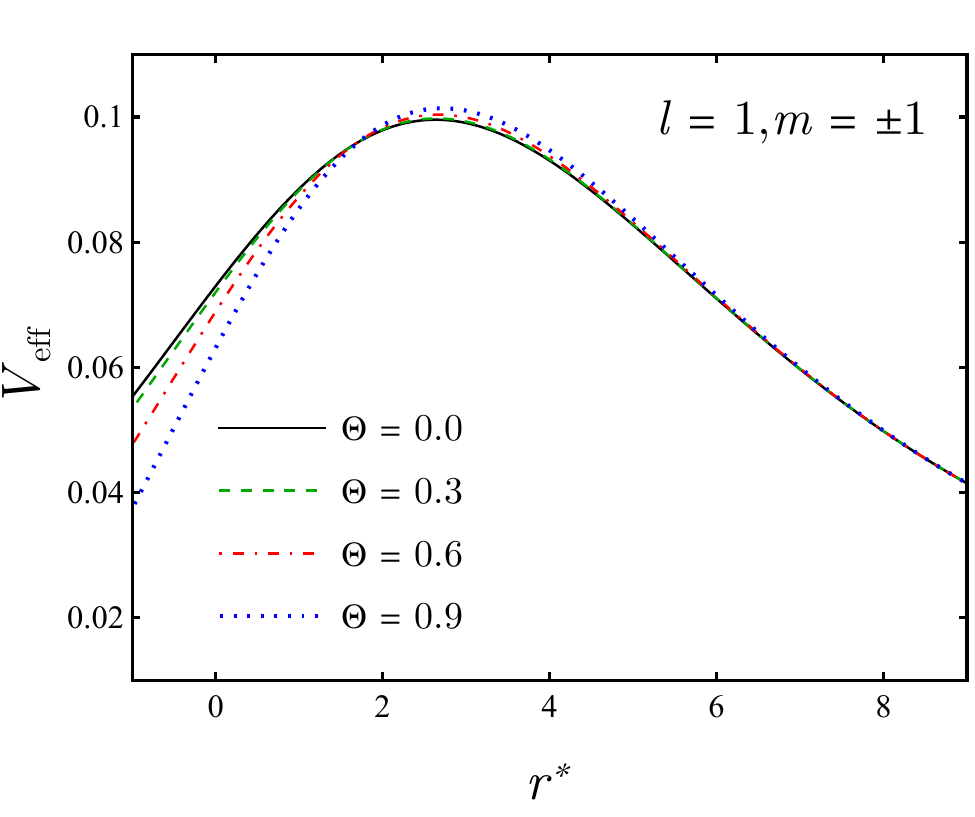} 
	\includegraphics[width=52mm]{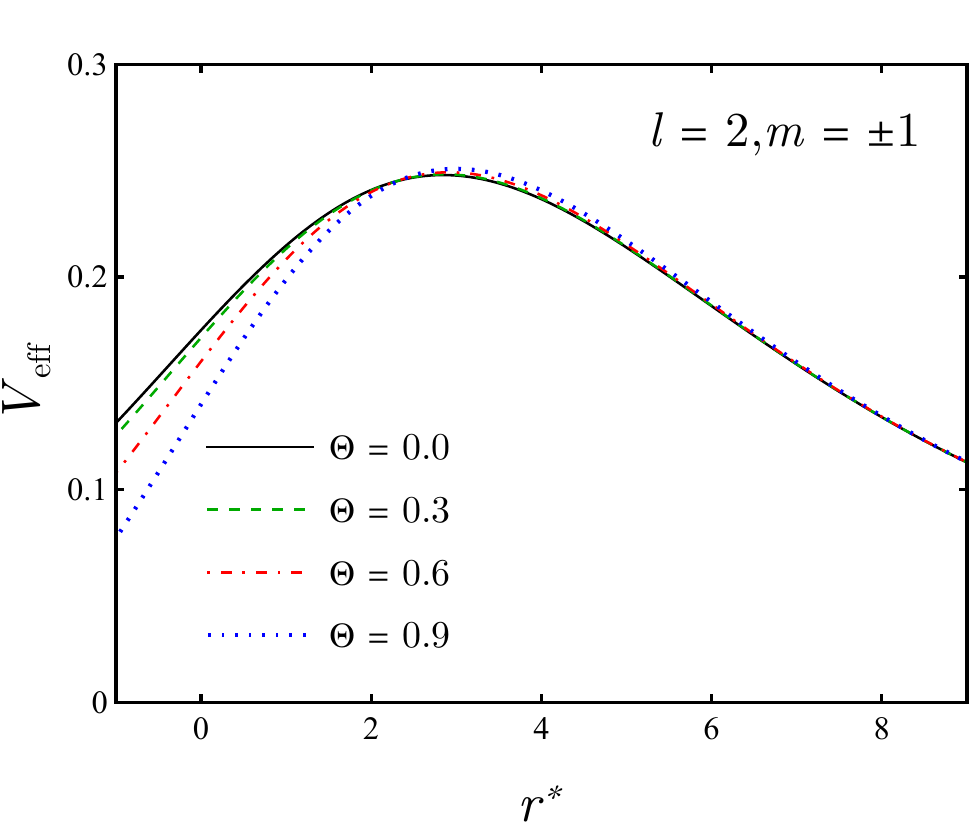}
    \includegraphics[width=52mm]{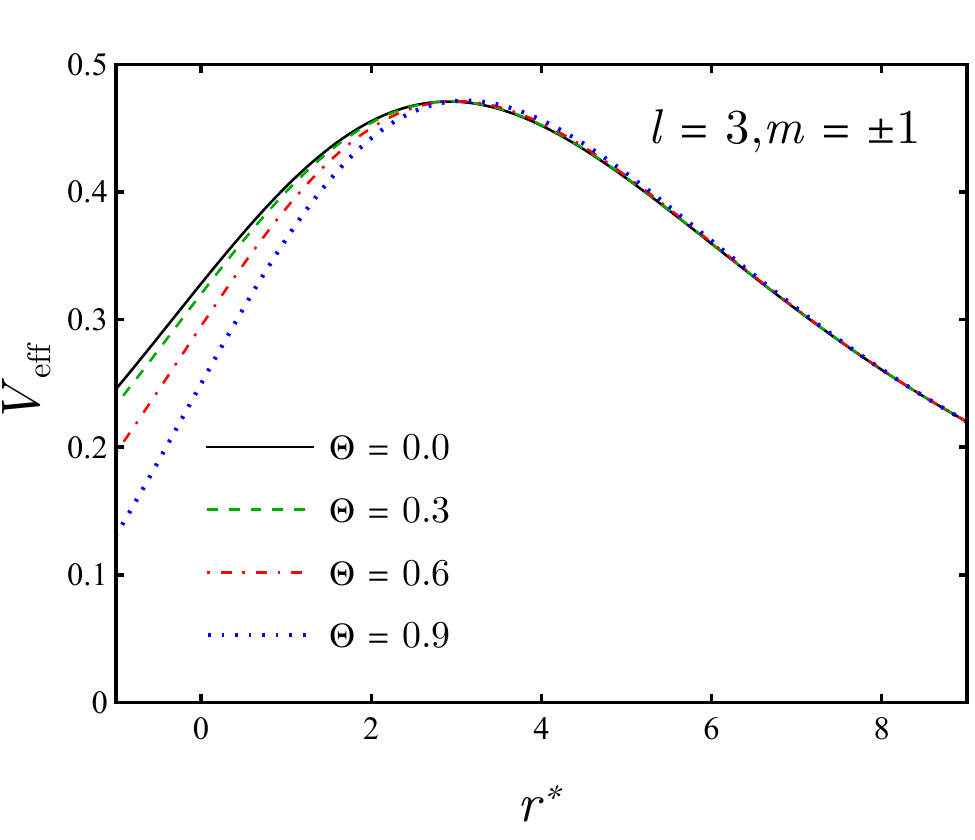}
	\caption{The effective potential concerning $r^*$ for different values of NC parameter when $M = 1$, $Q = 0.1$, $l = 1$, $2$, $3$ and $m = \pm 1$  }
		\label{fig:Veffrstar}
	\end{figure}

Fig. \ref{fig:Veffrstar} illustrates the effective potential, $V_{\text{eff}}$, corresponding to fixed values of the mass $M$, charge $Q = 0.1$, and different numbers of $l$ and $m$. Specifically, the left panel displays $V_{\text{eff}}$ for $l = 1$ ($m = \pm 1$), while the middle and right panels show the plots for $l = 2$ and $l = 3$ (with $m = \pm 1$), respectively.

The effect of the NC parameter on the effective potential is illustrated in Fig. \ref{fig:Veffrstar}. Variations in the value of \(\Theta\) lead to significant changes in the potential within NC spacetime. For all cases of $l$, an increase in the NC parameter leads to a higher peak in the potentials, indicating that the effective potential becomes a more substantial barrier to field transmission. This suggests that the NC modifications may significantly change the QNMs and other scattering properties of the scalar field. To explore these effects in more detail, we will utilize the effective potential to calculate the scattering characteristics of the scalar field in the following sections.


\subsection{Quasinormal modes}
In recent years, a variety of techniques have been employed to analyze quasinormal modes (QNMs), each offering distinct advantages and facing specific limitations \cite{iyer1987black,ferrari1984new,konoplya2002quasinormal,leaver1986spectral,konoplya2019higher,heidari2023investigation}. Among these, the Wentzel–Kramers–Brillouin (WKB) approximation has proven particularly Originally introduced by Schutz and Will \cite{schutz1985black} in the study of black hole scattering phenomena, this semi--analytical method was later extended by Iyer and Konoplya \cite{iyer1987black,konoplya2003quasinormal,konoplya2019higher}.

The application of the WKB method generally centers on the radial component $\mathcal{R_{lm}}(r)$ of the perturbation field. In the context of black hole spacetimes, the imposition of physically motivated boundary conditions plays a crucial role: at the event horizon, the field must consist solely of ingoing waves; conversely, at spatial infinity, only outgoing waves are permitted.

To compute QNMs, we adopt the third--order WKB formalism with the following equation
\begin{equation}\label{omegawkb}
\frac{{i(\omega _n^2 - V_{0})}}{{\sqrt { - 2V''_0} }} + \sum\limits_{j = 2}^3 {{\Omega _j} = n + \frac{1}{2}},
\end{equation}

In the WKB framework, $V_0$ denotes the peak of the effective potential, while $V_0''$ describes the second derivative with respect to the tortoise coordinate $r_*$ at that maximum. The terms $\Lambda_j(n)$ capture higher--order corrections associated with the $j$th--order WKB expansion \cite{konoplya2003quasinormal}. 

Table~\ref{Tab:AllQNMS} presents the QNM frequencies of scalar perturbations in a Reissner–Nordstr\"{o}m black hole with fixed mass 
$M = 1$ and charge $Q = 0.1$, for multiple values of the NC parameter. The results indicate a change in both the real and imaginary components of the QNMs as $\Theta$ increases. Specifically, the real part, associated with oscillation frequency, shows an enhancement, whereas the imaginary part, linked to damping, becomes more negative, signifying faster decay of perturbations. This trend is observed consistently across different multipole numbers $l = 1, 2$, azimuthal numbers $m$, and overtones $n$.

\begin{table}
    \centering
    \caption{QNMs of scalar perturbation of Reissner–Nordstr\"{o}m black hole for different $M = 1$, $Q = 0.1$. The multipole $l = 1 (m = \pm1)$ and $l =2 (m = \pm 1, 2)$ and corresponding overtones. \label{Tab:AllQNMS}}
\begin{tabular}{|cc|c|c|c|c|c|c|}
\hline
\multicolumn{2}{|c|}{} &
  $\Theta = 0.0$ &
  $\Theta = 0.1$ &
  $\Theta = 0.2$ &
  $\Theta = 0.3$ &
  $\Theta = 0.4$ \\ \hline
\multicolumn{1}{|c|}{\multirow{2}{*}{\begin{tabular}[c]{@{}c@{}}$l:1$,\\$m:\pm 1$\end{tabular}}} &
  $n:0$ &
 0.29162-0.09805i & 0.29172-0.09814i & 0.29204-0.09844i & 0.29255-0.09895i & 0.29323-0.09966i 
 \\ \cline{2-7} 
\multicolumn{1}{|c|}{} &
  $n:1$ &
  0.26277-0.30755i & 0.26307-0.30773i & 0.26395-0.30829i & 0.26537-0.30931i & 0.26722-0.31086i  \\ \hline
\multicolumn{1}{|c|}{\multirow{3}{*}{\begin{tabular}[c]{@{}c@{}}$l:2$ \\ $m:\pm 1$\end{tabular}}} &
  $n:0$ &
  0.48402-0.09685i & 0.48410-0.09697i & 0.48431-0.09731i & 0.48464-0.09789i & 0.48510-0.09868i   \\ \cline{2-7} 
\multicolumn{1}{|c|}{} &
  $n:1$ &
  0.46404-0.29595i & 0.46424-0.29627i & 0.46481-0.29722i & 0.46571-0.29881i & 0.46685-0.30106i    \\ \cline{2-7} 
\multicolumn{1}{|c|}{} &
  $n:2$ &
  0.43257-0.50366i & 0.43297-0.50411i & 0.43412-0.50547i & 0.43594-0.50780i & 0.43824-0.51116i    \\ \hline
\multicolumn{1}{|c|}{\multirow{3}{*}{\begin{tabular}[c]{@{}c@{}}$l:2$\\ $m:\pm2$\end{tabular}}} &
  $n:0$ &
  0.48402-0.09685i & 0.48412-0.09695i & 0.48440-0.09724i & 0.48485-0.09772i & 0.48548-0.09838i  \\ \cline{2-7} 
\multicolumn{1}{|c|}{} &
  $n:1$ &
  0.46404-0.29595i & 0.46424-0.29622i & 0.46482-0.29701i & 0.46574-0.29833i & 0.46694-0.30021i   \\ \cline{2-7} 
\multicolumn{1}{|c|}{} &
  $n:2$ &
  0.43257-0.50366i & 0.43293-0.50403i & 0.43399-0.50516i & 0.43568-0.50710i & 0.43785-0.50990i   \\ \hline
\end{tabular}
\end{table}

\begin{figure}[ht!]
	\centering
	\includegraphics[height=55mm]{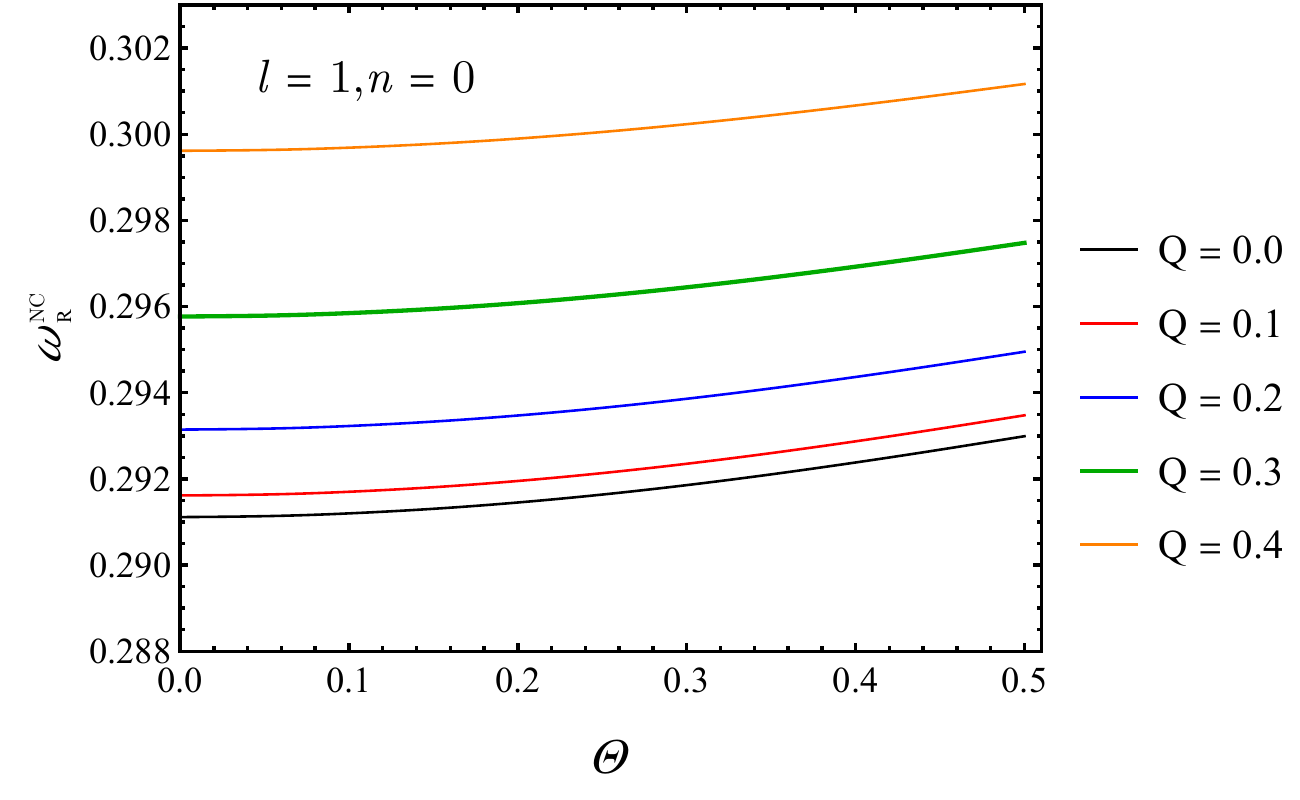}
	\includegraphics[height=55mm]{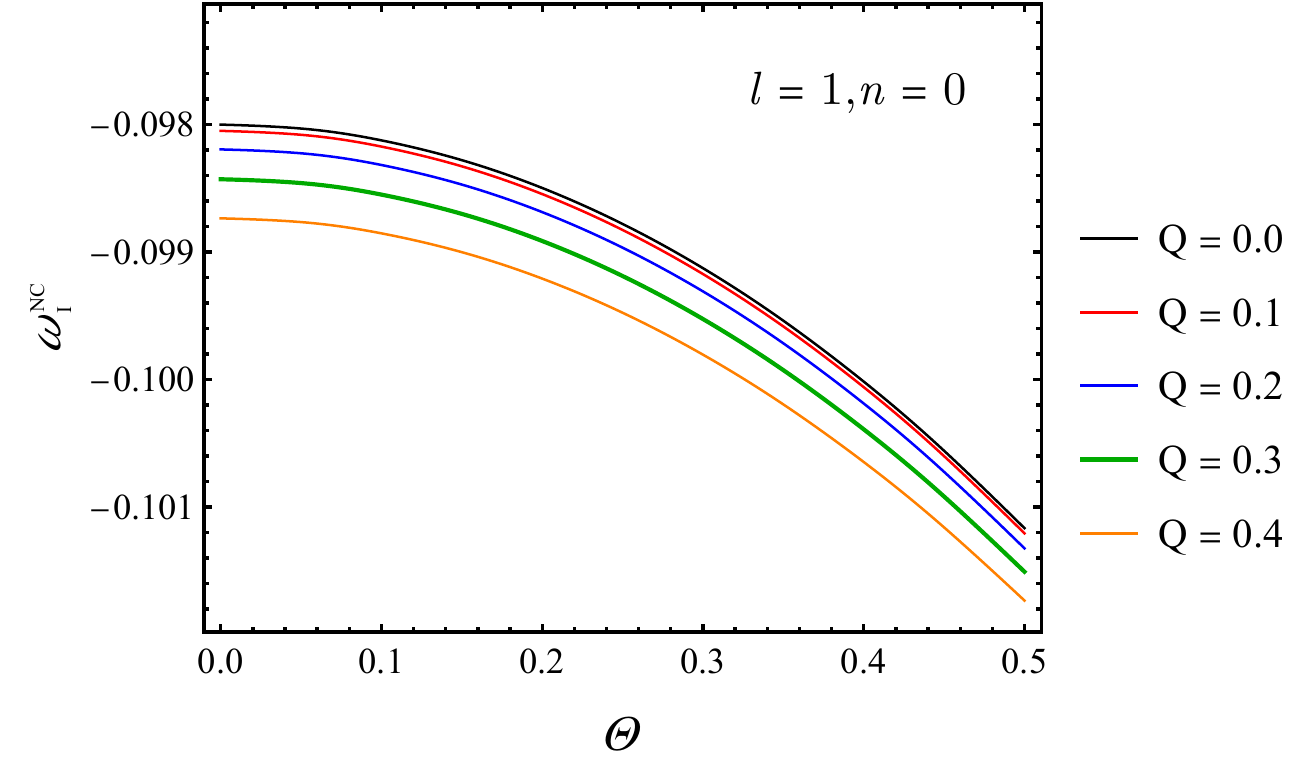}	
	\caption{The real andimaginary part of QNMs denoted as $\omega^{\text{NC}}_{\text{R}}$ and $\omega^{\text{NC}}_{\text{I}}$ are depicted concerning $\Theta$ for $M=1$, $l = 1 (m = \pm 1)$, $n = 0$ and different values of $Q$. }
	\label{fig:L1M1N0}
\end{figure}

\begin{figure}[ht!]
	\centering
	\includegraphics[width=80mm]{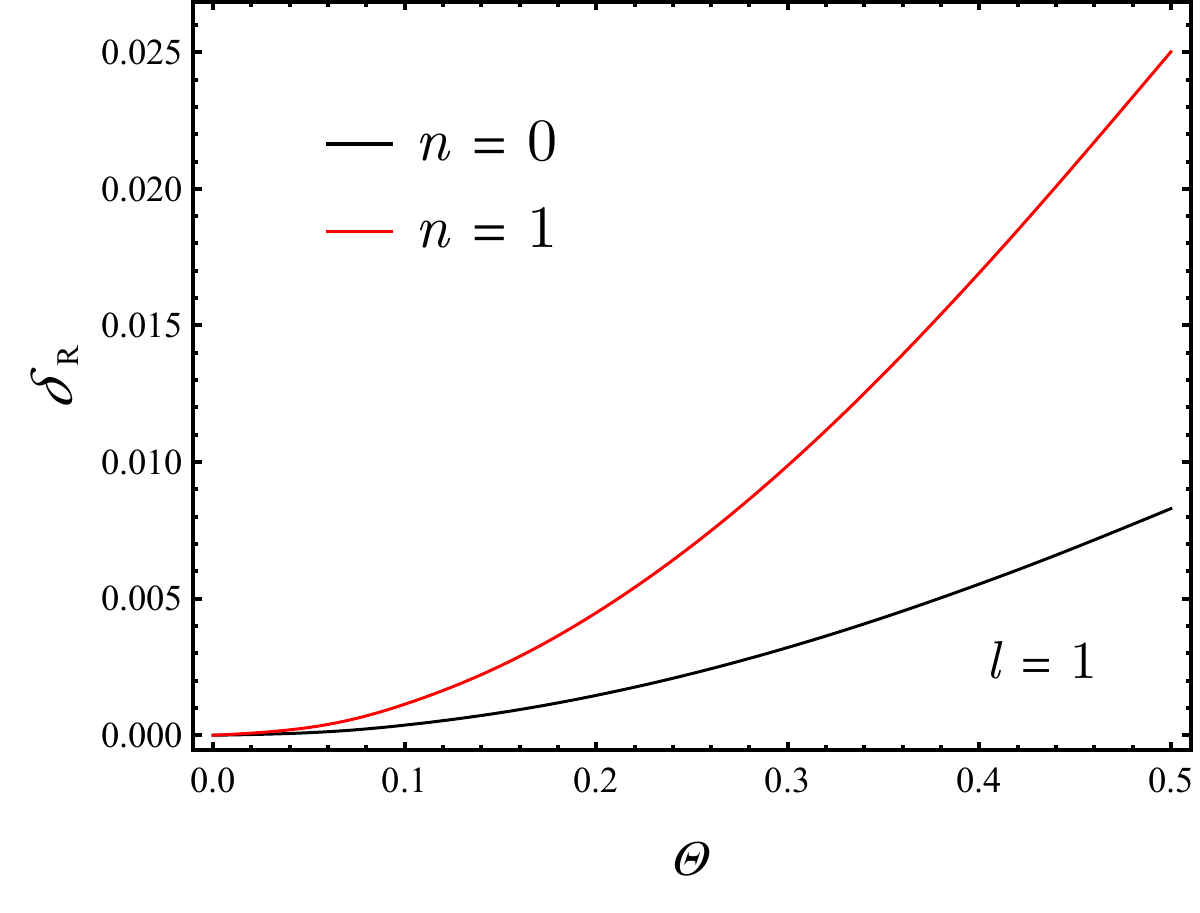}
	\includegraphics[width=80mm]{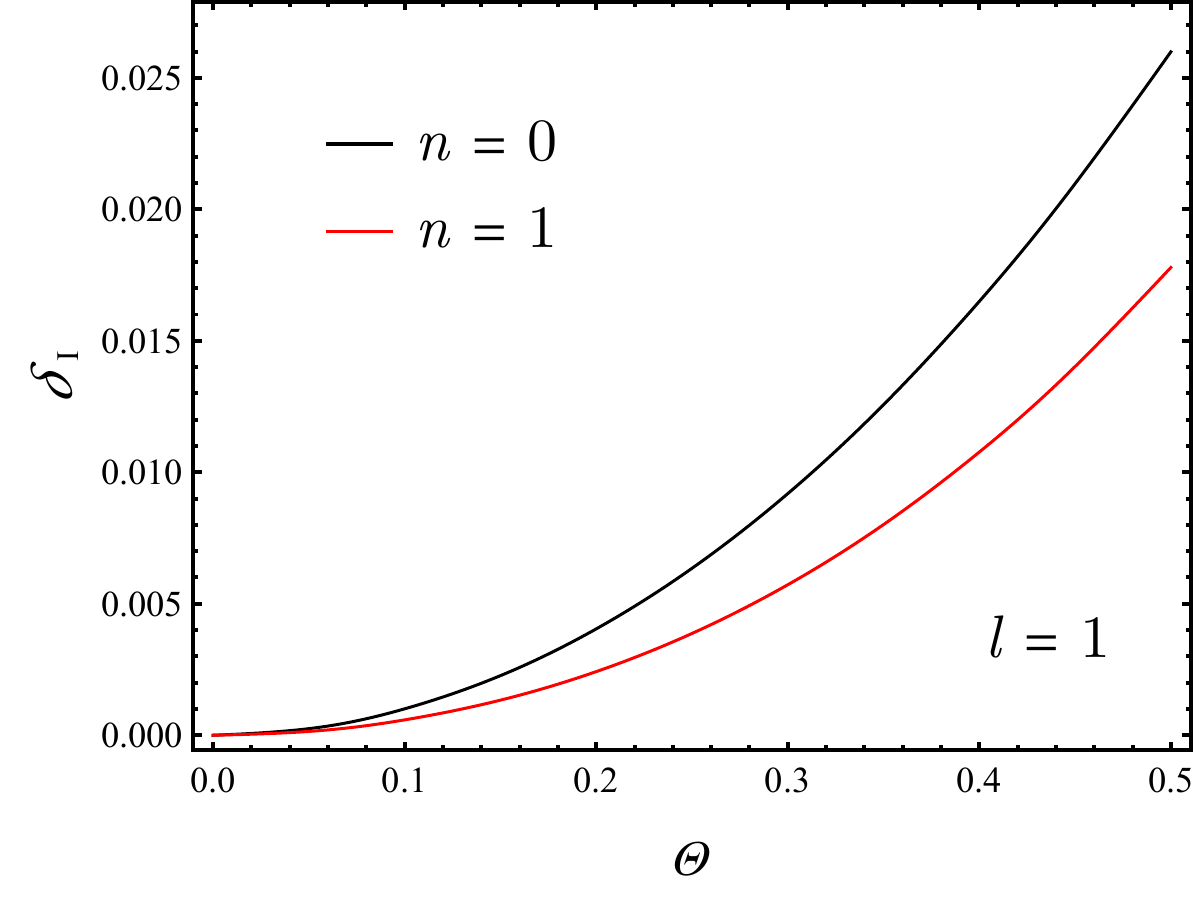}	
	\caption{Normalized deviations $\delta$ of the real and imaginary components of QNMs as functions of the NC parameter $\Theta$, for fixed values $M = 1$, $Q = 0.1$, $l = 1 (m = \pm 1)$, across different overtone numbers $n$. }
	\label{fig:L1M1N01}
\end{figure}

For better visualisation, the calculated QNMs for $l = 1$ are plotted and analyzed in Fig. \ref{fig:L1M1N0}--\ref{fig:L1M1N01}. The Fig. \ref{fig:L1M1N0} demonstrates the variation of the real and imaginary part of QNM frequencies with both the NC parameter $\Theta$ and charge $Q$ for $M = 1$, multipole number $l = 1 (m = \pm1)$ and overtone $n = 0$. There is a positive correlation between the real part of QNMs and $\Theta$, which means that NC effects enhance oscillation frequencies. At a fixed value of $\Theta$, a black hole with a higher charge $Q$ experiences a bigger propagation frequency as well.  On the other hand, the absolute value of the imaginary part, which determines the damping rate, increases with increasing $\Theta$. This indicates that higher NC impact enhances the dissipation of perturbations, causing them to decay faster. Moreover, for a given $\Theta$, an increase in the black hole's charge $Q$ results in a greater damping rate, as evidenced by a larger magnitude of the imaginary component $|\omega^{\text{NC}}_\text{I}|$.


To investigate the influence of overtone number, as it is so subtle, first we introduce a normalized deviation $\delta$ as 
    \begin{equation}
    \delta = \frac{\omega_{\scriptscriptstyle\rm {NC}}}{\omega_{\scriptscriptstyle\rm {RN}}}-1,
    \end{equation}
where ${\omega_{\scriptscriptstyle\rm {NC}}}$ and ${\omega_{\scriptscriptstyle\rm {RN}}}$ denote the QNMs of Reissner–Nordstr\"{o}m black hole in non--commutative and commutative black hole, respectively. Fig. \ref{fig:L1M1N01} represents the normalized deviation for both real and imaginary parts of QNMs for overtone $n = 0, 1$.

Our analysis shows that variations in the NC parameter produce qualitatively similar trends in both the real and imaginary components of QNMs across different overtone numbers. Notably, the normalized departure increases the real part more significantly for higher overtones, suggesting an enhancement in the oscillation frequency at higher modes. In contrast, the imaginary part, associated with the damping behavior, exhibits a stronger sensitivity to the NC parameter at lower overtone numbers, implying a shorter damping timescale for the black hole in the presence of NC spacetime.

\subsection{Comparison with other Approaches to QNMs}
There are other approaches to incorporate the non--comutativity in a charged black hole. We embed the non–commutative geometry through
a $(r, \theta)$ Moyal twist, applying the Seiberg–Witten map for Reissner–Nordstr\"{o}m black hole. In Ref.~\cite{ma2025quasinormal,yan2023shadows}, the hypothesis of a homogeneous charge distribution throughout the matter has been considered, and the spacetime metric for an electrically charged black hole in the framework of non--commutative geometry has been derived. In this non--commutative geometry framework, the mass and electric charge densities are modeled analogously using smeared distributions—either Gausthe sian $\rho^G_M = \frac{M}{(4\pi\Theta)^{3/2}} \, e^{-r^2/(4\Theta)}$, or Lorentzian $\rho^L_M = \frac{M\sqrt{\Theta}}{\pi^{3/2} (r^2 + \pi\Theta)^2}$. The charge density $\rho_Q$ is obtained directly from the mass density via the replacement $M \to Q$, reflecting the assumption of uniform charge-to-mass distribution.

To explore the influence of NC parameter on the QNM's behaviour in different approaches, we compared the general behaviour of the real and imaginary terms of QNMs. The summary of this comparison is represented in Tab.~\ref{tab:compQNM}, which reveals a consistent qualitative trend between our model and the Lorentzian smeared--source model: both the real oscillation frequency and the damping rate (magnitude of the imaginary part) increase with the NC parameter $\Theta$. This suggests a potentially universal stiffening effect on the spacetime's response to perturbations due to non--commutativity. Data for the Gaussian model in the cited references does not yield a clear or distinguishable trend, indicated by the dash in the table. This comparative analysis highlights how the specific implementation of NC geometry--whether through spacetime deformation or source smearing--can lead to observable and, in some cases, convergent predictions for dynamical signatures like QNMs.

\begin{table}[htbp]
\centering
\caption{Comparative trends of quasinormal mode frequencies with increasing non--commutative parameter $\Theta$ across different modeling frameworks. An upward arrow ($\uparrow$) indicates an increase in the respective quantity.}
\label{tab:compQNM}
\begin{tabular}{|@{}c|c|c|c|c@{}|}
\toprule
\textbf{NC Parameter} & \textbf{QNM} & \textbf{Our Work} & \textbf{Gaussian Dist.~\cite{yan2023shadows,ma2025quasinormal}} & \textbf{Lorentzian Dist.~\cite{ma2025quasinormal}} \\ 
\hline \hline
\multirow{2}{*}{$\Theta$ $\uparrow$} & $\omega_R$ & $\uparrow$ & -- & $\uparrow$ \\ 
\cline{2-5}
& $|\omega_I|$ & $\uparrow$ & -- & $\uparrow$ \\ \hline
\end{tabular}
\end{table}

\FloatBarrier


\section{Gravitational lensing}

Geodesics are fundamental to understanding the nature of spacetime, as they reveal its curvature and govern the motion of particles in gravitational fields. Exploring geodesics within the context of NC geometry has gained significant attention, as it examines the impact of quantum corrections on the spacetime structure. Exploring gravitational lensing in these contexts offers valuable insights into the behavior of particles at microscopic scales, where quantum effects are significant. Moreover, analyzing the geodesic structure of black hole spacetimes is essential for interpreting various astrophysical phenomena, including the characteristics of accretion disks or the formation of black hole shadows.

In this section, we comprehensively discuss the influence of the NC spacetime on the null--geodesic and gravitational lensing phenomena. The geodesic equation is derived by
\ie
\frac{\mathrm{d}^{2}x^{\mu}}{\mathrm{d}s^{2}} + \Gamma^\mu_{\nu \lambda} \frac{\mathrm{d}x^{\nu}}{\mathrm{d}s}\frac{\mathrm{d}x^{\lambda}}{\mathrm{d}s} = 0, \label{geodesicfull}
\fe
where $s$ and $\Gamma$ denote the affine parameter and the Christofell symbols, respectively
Examining how NC affects the paths of massless particles is our primary goal in the following discussion.

\subsection{Light trajectory}
\begin{figure}[ht!]
	\centering
	\includegraphics[width=80mm]{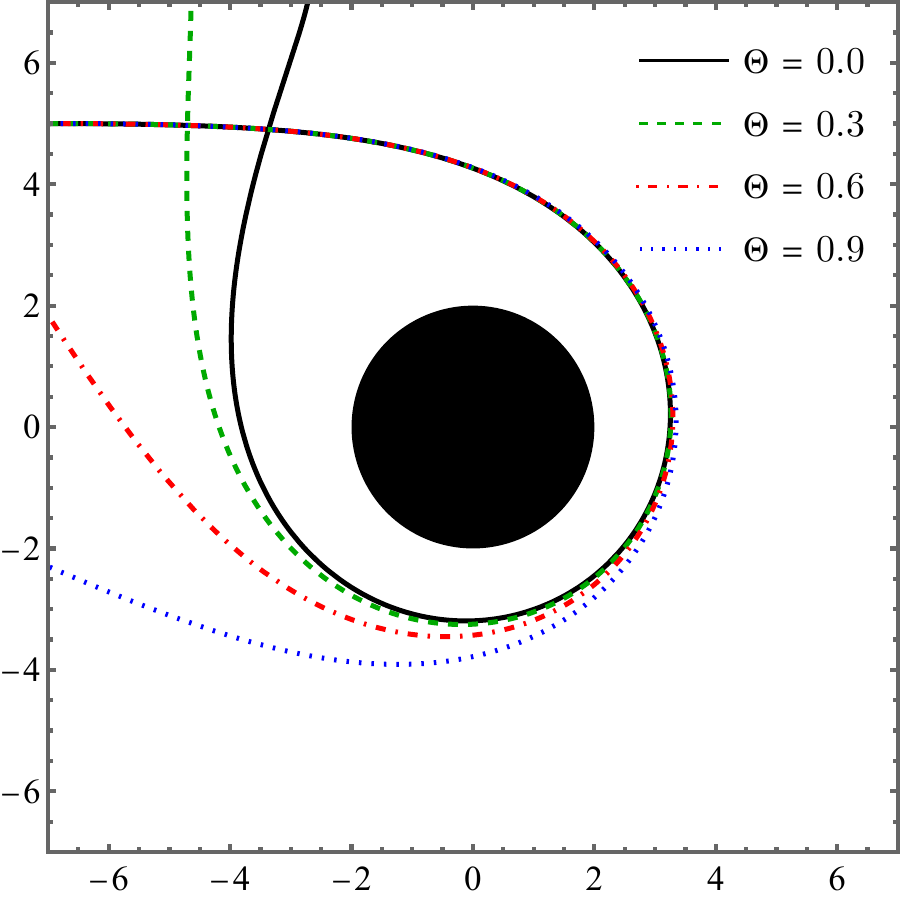}
	\caption{Light trajectory in the presence of non--commutative framework for a charged black hole with $2 M = 1$ and $Q = 0.1$ }
	\label{fig:light}
\end{figure}

To accomplish the trajectory of the light, a series of extended partial differential equations derived from Eq. \eqref{geodesicfull} must be determined. The statement mentioned above specifically produces four coupled partial differential equations that need to be resolved for each coordinate as

\begin{align} \label{geot}
    &\frac{\rm d t'}{\rm d s} = \Bigl( -2 r' t'{ \left(Q^2 r \tilde{B}+2 M r^2 \left(11 \Theta ^2 M+r^3-3 \Theta ^2 r\right)+42 \Theta ^2 Q^4\right)}\Bigl)/r\\ \nonumber
    &\Big(r^2 \left(-11 \Theta ^2 M^2-4 M r^3+4 \Theta ^2 M r+2 r^4\right)+Q^2 r \left(30 \Theta ^2 M+2 r^3-9 \Theta ^2 r\right)-14 \Theta ^2 Q^4\Big),\\ \label{geor}
    &\frac{\rm d r'}{\rm d s}  =-\Big(8 t'^2 \tilde{A}^2 \left[Q^2 r \tilde{B} +2 M r^2 \left(11 \Theta ^2 M+r^3-3 \Theta ^2 r\right)+42 \Theta ^2 Q^4\right]+r^4\left[Q^2 r^2 \tilde{C} \right.\\ \nonumber
    &+Q^4 r \tilde{D}+57 \Theta ^2 Q^6 -r^3 \left(64 \Theta ^2 M^3+64 M^2 \left(r^3-\Theta ^2 r\right)+M \left(15 \Theta ^2 r^2-64 r^4\right)\right.\\ \nonumber
    &\left.\left.+16 r^5\right)\right]\theta '^2+4 r^4 \sin ^2(\theta ) \left(\varphi '\right)^2 \left[Q^2 r^2 \tilde{F} -r^3 \left(2 \Theta ^2 M^3+2 M^2 \left(8 r^3-\Theta ^2 r\right)+M \left(\Theta ^2 r^2-16 r^4\right)\right.\right.\\ \nonumber
    &\left.\left.+4 r^5\right)-4 Q^4 r \left(4 \Theta ^2 M+r^3-2 \Theta ^2 r\right)+4 \Theta ^2 Q^6\right]+\frac{1}{\tilde{A}}\left[8r^4\left(2 Q^2 r^2 \left(-9 \Theta ^2 M^2\right.\right.\right.\\ \nonumber
    &\left.\left.-3 M r^3+10 \Theta ^2 M r+r^4-3 \Theta ^2 r^2\right)+Q^4 r \left(11 \Theta ^2 M+2 r^3-6 \Theta ^2 r\right)-2 \Theta ^2 Q^6\right]\\ \nonumber
    &\left.+M r^3 \left(6 \Theta ^2 M^2+4 M \left(r^3-2 \Theta ^2 r\right)-2 r^4+3 \Theta ^2 r^2\right)r'^2\right]\Big)\big/\\ \nonumber
    &\Big({8 r^5 \left(r^2 \left(3 \Theta ^2 M^2-2 M r \left(\Theta ^2+2 r^2\right)+2 r^4\right)+Q^2 r \left(-6 \Theta ^2 M+2 r^3+3 \Theta ^2 r\right)+2 \Theta ^2 Q^4\right)}\Big),\\ \label{geotheta}
    &\frac{\rm d \theta'}{\rm d s}  =\Big(\big[Q^2 r^2 \tilde{C}+Q^4 r \tilde{D}+57 \Theta ^2 Q^6-r^3 \left(64 \Theta ^2 M^3+64 M^2 \left(r^3-\Theta ^2 r+16 r^5\right)\right.\\ \nonumber
    &\left.\left.+M \left(15 \Theta ^2 r^2-64 r^4\right)\right)\right)\big]4r'\theta '+r \sin 2\theta  \varphi '^2 \tilde{A}\big[16 \Theta ^2 Q^4+r^2 \tilde{E} \\ \nonumber
    & +Q^2 r \left(-32 \Theta ^2 M+16 r^3+15 \Theta ^2 r\right)\big]\Big)\big/\Big(2 r \tilde{A} \left(r^2 \left(64 \Theta ^2 M^2-32 M r \left(\Theta ^2+r^2\right)\right)\right.\\ \nonumber
    &\left.+r^2 \left(\Theta ^2+16 r^2\right)+2 Q^2 r \left(-72 \Theta ^2 M+8 r^3+27 \Theta ^2 r\right)+57 \Theta ^2 Q^4\right)\Big),\\ \label{geophi}
 &\frac{\rm d \varphi'}{\rm d s}  =-2 \sin \theta \varphi ' \Big(r \theta ' \cos \theta \tilde{A} \left[r^2 \tilde{E}+Q^2 r \tilde{G} +16 \Theta ^2 Q^4\right] \\ \nonumber
 &-4 \sin \theta  r'\left[Q^2 r^2 \tilde{F} -r^3 \left(2 \Theta ^2 M^3+2 M^2 \left(8 r^3-\Theta ^2 r\right)+M \left(\Theta ^2 r^2-16 r^4\right)+4 r^5\right)\right.\\ \nonumber
 &\left.-4 Q^4 r \left(4 \Theta ^2 M+r^3-2 \Theta ^2 r\right)+4 \Theta ^2 Q^6\right]\Big)/\Big(r \tilde{A}\big(4 \sin ^2\theta \left[r^2 \left(2 \Theta ^2 M^2-8 M r^3\right.\right.\\ \nonumber
 &\left.\left.-4 \Theta ^2 M r+4 r^4+\Theta ^2 r^2\right)+Q^2 r \left(-8 \Theta ^2 M+4 r^3+5 \Theta ^2 r\right)+4 \Theta ^2 Q^4\right]+5 \Theta ^2 r^2 \cos ^2\theta  \tilde{A}\big)\Big),
\end{align}
where the terms  $\tilde{A}$--$\tilde{H}$ are presented in the appendix II. The light trajectories for a black hole with $2M = 1$, $Q = 0.1$ and various values of the NC parameter are demonstrated in Fig. \ref{fig:light}. It shows that the light trajectory for greater values of $\Theta$ is deflected less in the vicinity of the black hole. Therefore, the NC framework diminishes the gravitational lensing effects of the black hole on light rays. 
This effect of NC on the charged black hole is consistent with the finding reported in Ref. \cite{heidari2024exploring}, where the NC framework in the Schwarzschild black hole disempowers the gravitational lensing effect.

\subsection{Photonic Radius}
The study of photonic and shadow radius of a back hole represents a crucial area of research \cite{zeng2022shadows,panah2025some,heidari2025gravitational,HAMIL2023101293,anacleto2023absorption,yan2023shadows}. Interest in this topic has grown significantly, particularly after the Event Horizon Telescope (EHT) captured the image of the $Sgr A^*$ and $M87^*$ black hole \cite{ball2019first,gralla2021can,akiyama2019first}. To analyze the shadow radius and the behavior of null geodesics, as outlined in Ref. \cite{batic2015light}, we employ a diagonal metric with parameters \( g^{\scriptscriptstyle {(\Theta},\scriptscriptstyle{Q)}}_{ij} \), expressed as follows

\begin{equation} 
{g^{\scriptscriptstyle {(\Theta},\scriptscriptstyle{Q)}}_{\mu \nu }}\mathrm{d}{x^\mu }\mathrm{d}{x^\nu } =  - g^{\scriptscriptstyle {(\Theta},\scriptscriptstyle{Q)}}_{tt}\mathrm{d}{t^2} + g^{\scriptscriptstyle {(\Theta},\scriptscriptstyle{Q)}}_{rr}\mathrm{d}{r^2} +g^{\scriptscriptstyle {(\Theta},\scriptscriptstyle{Q)}}_{\theta \theta}\mathrm{d}{\theta ^2} + g^{\scriptscriptstyle {(\Theta},\scriptscriptstyle{Q)}}_{\phi \phi}{{\mathop{\rm \sin}\nolimits} ^2}\theta \mathrm{d}{\varphi ^2}.\\
\end{equation}

By applying this metric to the Lagrangian $\mathcal{L}$ formulation, defined as  
\begin{equation}
2\mathcal{L} = - g^{\scriptscriptstyle {(\Theta},\scriptscriptstyle{Q)}}_{tt}{{\dot t}^2} + g^{\scriptscriptstyle {(\Theta},\scriptscriptstyle{Q)}}_{rr}{{\dot r}^2} +g^{\scriptscriptstyle {(\Theta},\scriptscriptstyle{Q)}}_{\theta \theta}{{\dot \theta }^2} + g^{\scriptscriptstyle {(\Theta},\scriptscriptstyle{Q)}}_{\phi \phi}{{\dot \varphi }^2}.
\end{equation}

Assuming geodesic motion confined to the equatorial plane, $\theta = \frac{\pi }{2}$, we set $\dot \theta  = 0$. The corresponding Euler--Lagrange equations for coordinates $t$ and $\varphi$
results in two conserved quantities, the energy and angular momentum, denoted by $E$ and $L$, as

\begin{equation}\label{constant}
E = g^{\scriptscriptstyle {(\Theta},\scriptscriptstyle{Q)}}_{tt}\dot t \quad\mathrm{and}\quad L = g^{\scriptscriptstyle {(\Theta},\scriptscriptstyle{Q)}}_{\phi \phi}\dot \varphi .
\end{equation}

Defining the impact parameter as $b=\frac{L}{E}$, we analyze the trajectory of light, which follows the condition $\mathcal{L} =0$. This leads to the equation  

\begin{equation}\label{light}
- g^{\scriptscriptstyle {(\Theta},\scriptscriptstyle{Q)}}_{tt}{{\dot t}^2} + g^{\scriptscriptstyle {(\Theta},\scriptscriptstyle{Q)}}_{rr}{{\dot r}^2} + g^{\scriptscriptstyle {(\Theta},\scriptscriptstyle{Q)}}_{\phi \phi}{{\dot \varphi }^2} = 0 .
\end{equation}

By substituting Eq. \eqref{constant} into Eq. \eqref{light}, the light trajectory in the equatorial plane is obtained as

\begin{equation}
{\left(\frac{{\mathrm{d}r}}{{\mathrm{d}\varphi }}\right)^2} + {\mathcal{V}_{\text{eff}}} = 0 , \quad\quad \text{where}\quad\quad {\mathcal{V}_{\text{eff}}} =\frac{g^{\scriptscriptstyle {(\Theta},\scriptscriptstyle{Q)}}_{\phi \phi}}{g^{\scriptscriptstyle {(\Theta},\scriptscriptstyle{Q)}}_{rr}}\left(\frac{1}{b^2} \frac{g^{\scriptscriptstyle {(\Theta},\scriptscriptstyle{Q)}}_{\phi \phi}}{g^{\scriptscriptstyle {(\Theta},\scriptscriptstyle{Q)}}_{tt}}- 1\right) .
\end{equation}
The conditions for circular photon orbits are given by  

\begin{equation}
{\mathcal{V}_{\text{eff}}} = \frac{{\mathrm{d}{\mathcal{V}_{\text{eff}}}}}{{\mathrm{d}r}} = 0 .
\end{equation}

Solving this equation provides the radius of the photon sphere $r_{\text{ph}}$, determined by  

\begin{equation}
g^{\scriptscriptstyle {(\Theta},\scriptscriptstyle{Q)}}_{tt}{}'g^{\scriptscriptstyle {(\Theta},\scriptscriptstyle{Q)}}_{\phi \phi} - g^{\scriptscriptstyle {(\Theta},\scriptscriptstyle{Q)}}_{\phi \phi}{}'g^{\scriptscriptstyle {(\Theta},\scriptscriptstyle{Q)}}_{tt}= 0.
\end{equation}
where prime denotes the derivative to the radius. Based on the metric functions described in Eq. \eqref{metric1}, and through further algebraic manipulations, the photon radius is determined as  

\begin{align}\nonumber
    &\frac{\Theta ^2 }{2 r_\text{ph}^5 \left(r_\text{ph} (r_\text{ph}-2 M)+Q^2\right)}\Big[Q^2 r_\text{ph}^2 \left(-492 M^2+344 M r_\text{ph}-57 r_\text{ph}^2\right)+Q^4 r_\text{ph} (438 M-169 r_\text{ph})\\ 
&+4 M r_\text{ph}^3 (3 M-r_\text{ph}) (11 M-5 r_\text{ph})-112 Q^6]+\frac{2 \left({r_{\text{ph}}} ({r_{\text{ph}}}-3 M)+2 Q^2\right)}{r_{\text{ph}}}=0
\end{align}

The values of $r_{\text{ph}}$ are computed for $M=1$ and varying $Q$ and $\Theta$ to examine the effect of NC on photon spheres. The data in Table \ref{Tab:rphoton} reveal the dependence of the photon sphere radius on the charge and the NC parameter. As the charge $Q$ increases from 0.1 to 0.4, the photon sphere radius systematically decreases for all values of $\Theta$. This trend suggests that higher charge values lead to a more compact photon sphere. Additionally, for a fixed charge, increasing the NC parameter also results in a gradual decrease in $r_\text{ph}$. Our findings align with results from the Gaussian smeared--source model in Ref. \cite{yan2023shadows}, which reports a similar decrease in $r_{\text{ph}}$ with increasing $Q$ and $\Theta$. This consistent trend across different non--commutative implementations suggests a robust, model-independent contraction of the photon sphere due to both charge and NC parameter, which potentially affects observable astrophysical phenomena we are interested in exploring.

\begin{table}[!ht]
    \centering
    \caption{The photon sphere radius $r_{\text{ph}}$ for different values of the parameter $\Theta$ and charge $Q$.}
    \begin{tabular}{|c|c|c|c|c|}
    \hline
        \textbf{$\Theta$} & \textbf{Q = 0.1} & \textbf{Q = 0.2} & \textbf{Q = 0.3} & \textbf{Q = 0.4} \\ \hline \hline

 \textbf{0.0} & 2.99332 & 2.97309 & 2.93875 & 2.88924 \\ \hline
 \textbf{0.1} & 2.99318 & 2.97295 & 2.93861 & 2.88911 \\ \hline
 \textbf{0.2} & 2.99276 & 2.97254 & 2.93821 & 2.88871 \\ \hline
 \textbf{0.3} & 2.99207 & 2.97185 & 2.93753 & 2.88805 \\ \hline
 \textbf{0.4} & 2.9911 & 2.97089 & 2.93658 & 2.88712 \\ \hline
 \textbf{0.5} & 2.98985 & 2.96965 & 2.93535 & 2.88592 \\ \hline
 \textbf{0.6} & 2.98833 & 2.96813 & 2.93385 & 2.88445 \\ \hline
 \textbf{0.7} & 2.98652 & 2.96634 & 2.93208 & 2.88271 \\ \hline
 \textbf{0.8} & 2.98444 & 2.96427 & 2.93002 & 2.88069 \\ \hline
    \end{tabular}
    \label{Tab:rphoton}
\end{table}


\subsubsection{Topological features of the photonicyieldsre}
\begin{figure}[ht!]
	\centering
    \includegraphics[width=80mm]{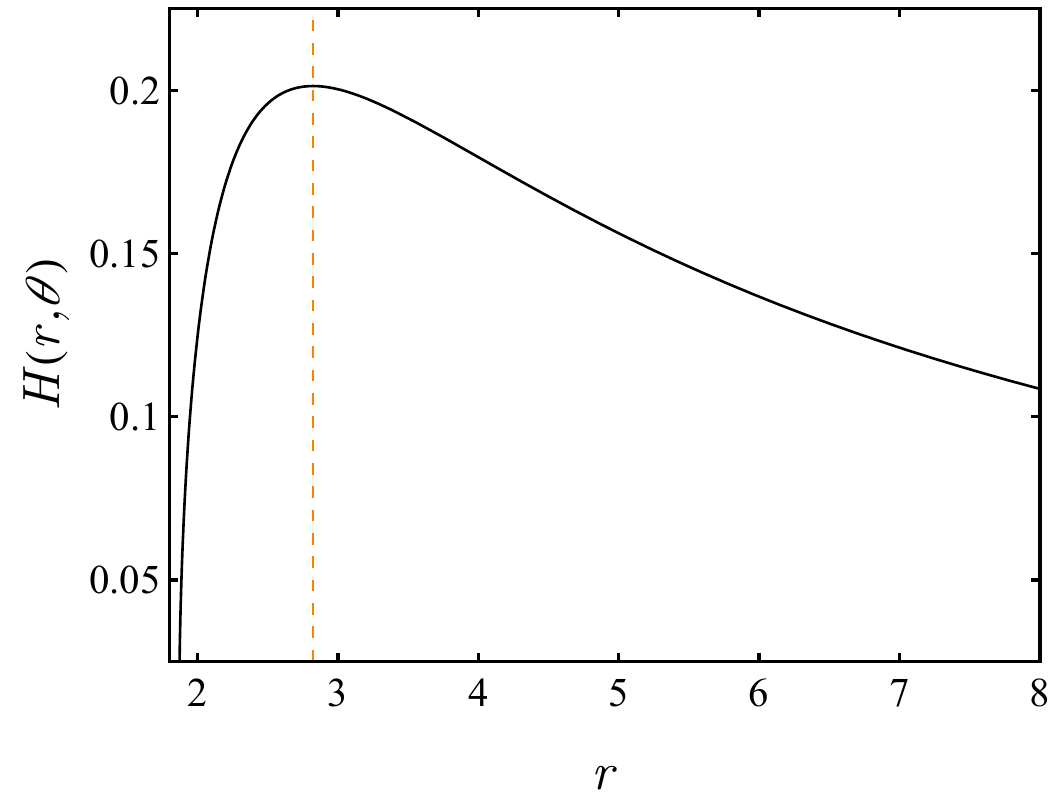}
	\caption{The behaviour of the topological potential $H(r,\theta)$ is shown against the $r$. The photonic radius is indicated by a red dashed line.}
	\label{fig:Hr}
\end{figure}

The photonic radius can be classified as stable or unstable. In standard spherical symmetric hole \ cite es, the stable circular orbits type yield instability in spacetime. Additionally, the unstable type can determine the shadows of the black hole\cite{Vec0-Wei2020,Vec5-Cunha2020,Vec7-Sadeghi2024,Vec3-BahrozBrzo2025,Vec9-alipour2024weak,Vec1-Sadeghi2023,Vec6-Pantig2025}. In this part, the stability of the photon sphere is explored based on the topological method \cite{Vec0-Wei2020}. First, a regular potential function is introduced as 
\begin{align}\label{eq:Hr}
&H(r, \theta) =  \sqrt{\frac{-g^{\scriptscriptstyle {(\Theta},\scriptscriptstyle{Q)}}_{tt}}{g^{\scriptscriptstyle {(\Theta},\scriptscriptstyle{Q)}}_{\phi\phi}}}=\\ \nonumber
&\Bigg(8\Gamma  \Big[\Theta ^2 \left(3 Q^2 r (3 r-10 M)+M r^2 (11 M-4 r)+14 Q^4\right)-2 \Gamma  r^4\Big]/\\ \nonumber
        &\Big[4 r^4 \sin ^2(\theta ) \left(\Theta ^2 \left(r^2 \left(2 M^2-4 M r+r^2\right)+Q^2 r (5 r-8 M)+4 Q^4\right)+4 \Gamma  r^4\right)+5 \Gamma  \Theta ^2 r^6 \cos ^2(\theta )\Big]\Bigg)^{1/2}.
\end{align}

where $\Gamma=r (r-2 M)+Q^2$. In the above expression,  "$\ sin\theta$" and "$\cos\theta$" serve as a factor that allows systematic topological analysis. Fig. \ref{fig:Hr} shows the behavior of potential $H(r,\theta)$. The photon sphere locations correspond to critical points satisfying $\partial_r H = 0$. The maximum of the potential, for $M = 1$, $Q = 0.5$, and $\Theta = 0.1$, is located at $r_{\text{ph}}= 2.82288$ and demonstrates that this photonic position is unstable. Minor disturbances can destabilize the photon trajectory, as seen in this plot, which can lead to the particle escaping outward or being caught by the black hole.

We define an associated vector field $\bm{\varphi}$ with the following components 
\begin{align}
\varphi_ r &= \sqrt{g^{\scriptscriptstyle {(\Theta},\scriptscriptstyle{Q)}}_{tt}} \, \partial_r H(r, \theta), \label{eq:phi_r} \\
\varphi_\theta &= \frac{1}{\sqrt{{g^{\scriptscriptstyle {(\Theta},\scriptscriptstyle{Q)}}_{\theta \theta}}}}\, \partial_\theta H(r, \theta), \label{eq:phi_theta}
\end{align}
whose zeros correspond to the photon sphere positions. The complex representation $ \varphi = \varphi_r +i \varphi_\theta$ facilitates further topological analysis.
The normalized vector components are expressed as

\begin{equation}
n_r = \frac{\varphi_r}{\|\bm{\varphi}\|}, \quad n_\theta = \frac{\varphi_\theta}{\|\bm{\varphi}\|}, \quad \|\bm{\varphi}\| = \sqrt{\varphi_r^2 + \varphi_\theta^2}
\label{eq:normalized}
\end{equation}

The photonic radius can be analyzed as topological defects occurring at points where the component of the field $\phi$ vanishes. When a closed loop encloses such a zero point, it signifies that the net topological charge corresponds to the winding number. Each photon sphere associated with a black hole corresponds to a unique topological charge, denoted by $\mathcal{Q}$, which takes the discrete values of $\pm1$.
Following Ref. \cite {Vec5-Cunha2020,Vec3-Duane1984}, the photonic sphere with $\mathcal{Q}=-1$ is unstable, and the one with $\mathcal{Q}=+1$ corresponds to a stable photonic radius.

\begin{figure}[ht!]
	\centering
	\includegraphics[width=80mm]{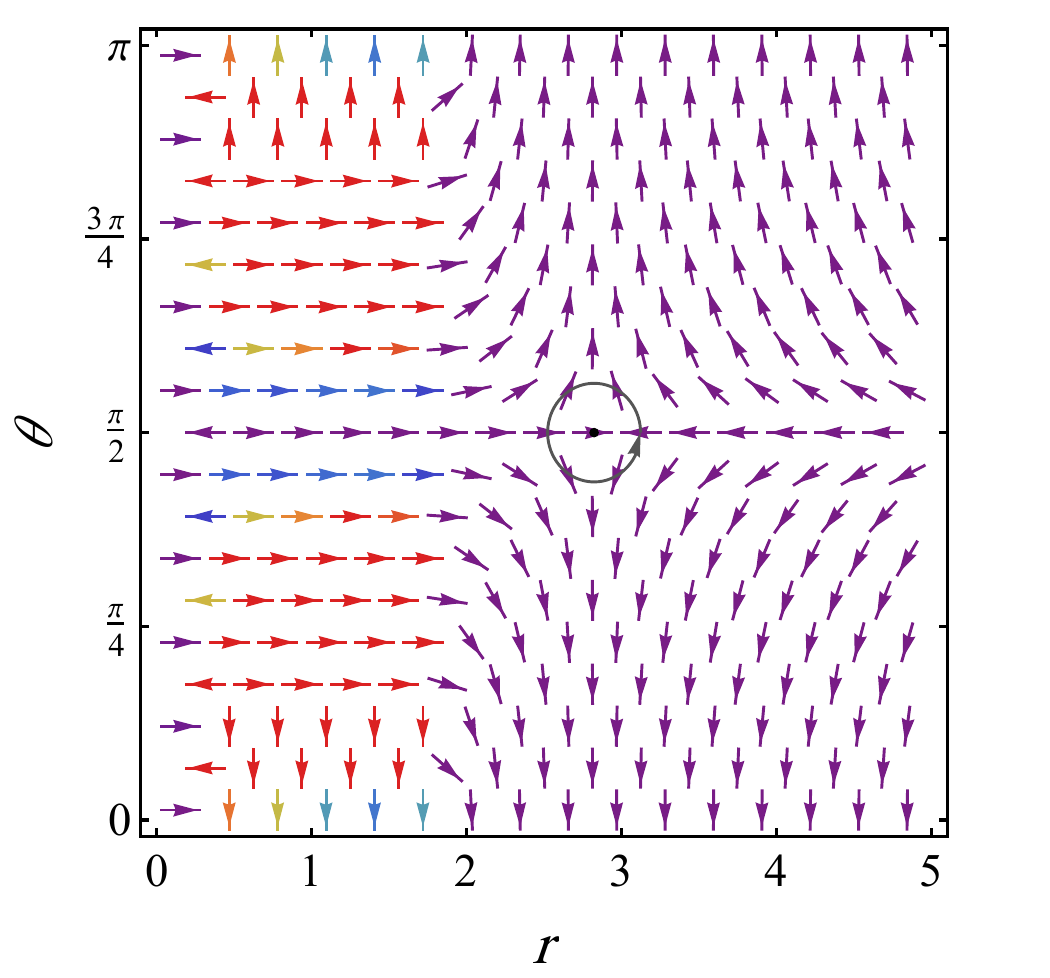}
    \caption{The normalized vector field on the $(r, \theta)$ plane for $M = 1$, $Q = 0.5$ and $\Theta = 0.1$. The black loop encircles theonic sphere point $(2.82288,\frac{\pi}{2})$. }
	\label{fig:vector}
\end{figure}

Fig. \ref{fig:vector} illustrates the structure of the vector field, highlighting the only photon sphere larger than the event horizon located at $r_{\text{ph}}= 2.95871$, where the field lines incorporate with a topological charge of $-1$. Following the proposed classification proposed in Ref . \cite{Vec0-Wei2020, Vec5-Cunha2020, Vec7-Sadeghi2024}, this photon sphere remains intrinsically unstable.


\subsubsection{Stability of the photon sphere}

The stability of photon spheres near black holes is primarily influenced by the geometric and topological features of the associated optical spacetime, with conjugate points playing a pivotal role. Under perturbations, the behavior of photon trajectories is determined by the stability properties of the photon sphere. For unstable configurations, even small deviations can cause photons to either fall into the black hole or escape to infinity. In contrast, stable photon spheres allow photons to remain confined in nearby bounded orbits \cite{qiao2022curvatures,qiao2022geometric}.

The existence or absence of conjugate points within the spacetime manifold is a key factor influencing the stability of photon trajectories. Stable photon spheres are characterized by the presence of conjugate points, while their unstable counterparts are devoid of them. The Cartan--Hadamard theorem establishes a relationship between the Gaussian curvature \( \mathcal{K}(r) \) and the occurrence of conjugate points, offering a framework to evaluate the stability of critical orbits \cite{qiao2024existence}. In this context, the null geodesics—defined by the condition \( \mathrm{d}s^2 = 0 \) -- can be represented as follows \cite{araujo2024effects,heidari2024absorption}
\ie
\mathrm{d}t^2=\gamma_{ij}\mathrm{d}x^i \mathrm{d}x^j = -\frac{g^{\scriptscriptstyle {(\Theta},\scriptscriptstyle{Q)}}_{rr}(r)}{g^{\scriptscriptstyle {(\Theta},\scriptscriptstyle{Q)}}_{tt}(r)}\mathrm{d}r^2  -\frac{\Bar{g}^{\scriptscriptstyle {(\Theta},\scriptscriptstyle{Q)}}_{\varphi\varphi}(r)}{g^{\scriptscriptstyle {(\Theta},\scriptscriptstyle{Q)}}_{tt}(r)}\mathrm{d}\varphi^2   ,
\fe
where \( i \) and \( j \) range from \( 1 \) to \( 3 \), \( \gamma_{ij} \) denotes the components of the optical metric, and
$\Bar{g}^{\scriptscriptstyle {(\Theta},\scriptscriptstyle{Q)}}_{\varphi\varphi} (r) \equiv g^{\scriptscriptstyle {(\Theta},\scriptscriptstyle{Q)}}_{\varphi\varphi}(r,\theta = \pi/2)$. Moreover, the Gaussian curvature is calculated by \cite{qiao2024existence}
\ie
\label{dffdsf}
\mathcal{K}(r,\Theta,Q) = \frac{\mathcal{R}}{2} =  \frac{g^{\scriptscriptstyle {(\Theta},\scriptscriptstyle{Q)}}_{tt}(r)}{\sqrt{g^{\scriptscriptstyle {(\Theta},\scriptscriptstyle{Q)}}_{rr}(r) \,  \Bar{g}^{\scriptscriptstyle {(\Theta},\scriptscriptstyle{Q)}}_{\varphi\varphi}(r)}}  \frac{\partial}{\partial r} \left[  \frac{g^{\scriptscriptstyle {(\Theta},\scriptscriptstyle{Q)}}_{tt}(r)}{2 \sqrt{g^{\scriptscriptstyle {(\Theta},\scriptscriptstyle{Q)}}_{rr}(r) \, \Bar{g}^{\scriptscriptstyle {(\Theta},\scriptscriptstyle{Q)}}_{\varphi\varphi}(r) }}   \frac{\partial}{\partial r} \left(   \frac{\Bar{g}^{\scriptscriptstyle {(\Theta},\scriptscriptstyle{Q)}}_{\varphi\varphi}(r)}{g^{\scriptscriptstyle {(\Theta},\scriptscriptstyle{Q)}}_{tt}(r)}    \right)    \right],
\fe
where $\mathcal{R}$ represents the Ricci scalar in two dimensions \cite{araujo2024effects}. When \( \Theta \) is sufficiently small, an explicit approximation of the Gaussian curvature is given by

\ie
\begin{split}
\label{gaussiancurvature}
\mathcal{K}(r,\Theta,Q) & =  \frac{3 M^2-2 M r}{r^4}+\frac{3 Q^2 (r-2 M)}{r^5} \\
& + \Theta ^2 \left\{\frac{-624 M^4+848 M^3 r-374 M^2 r^2+54 M r^3-r^4}{4 r^7 (r-2 M)}\right. \\
& \left. +\frac{Q^2 \left(-2964 M^4+5224 M^3 r-3374 M^2 r^2+934 M r^3-91 r^4\right)}{2 r^8 (r-2 M)^2}\right\}.
\end{split}
\fe

\begin{figure}
    \centering
    \includegraphics[scale=0.65]{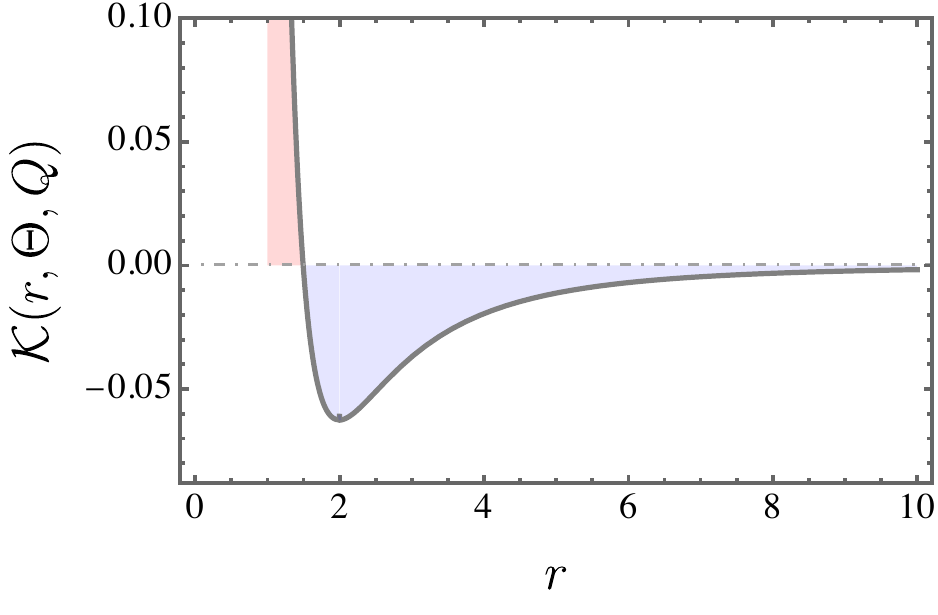}
    \caption{The Gaussian curvature $\mathcal{K}(r,\Theta, Q)$  is depicted, with clear distinction between the regions corresponding to stable (pink) and unstable (purple) configurations. The analysis is carried out using the parameter values \( M = 1 \), \( Q = 0.1 \), and \( \Theta = 0.01 \).
}
    \label{gauss}
\end{figure}

As discussed in Refs .~\cite {qiao2022curvatures, qiao2022geometric, qiao2024existence}, the stability of photon spheres can be inferred from the sign of the Gaussian curvature \( \mathcal{K}(r,\Theta, Q) \). Specifically, a negative curvature (\( \mathcal{K}(r,\Theta,Q) < 0 \)) signifies instability, whereas a positive curvature (\( \mathcal{K}(r,\Theta,Q) > 0 \)) indicates stability. To visualize this behavior, Fig.~\ref{gauss} presents \( \mathcal{K}(r,\Theta, Q) \) as a function of the radial coordinate \( r \), outlining the regions corresponding to stable and unstable photon spheres. The analysis is performed for \( M = 1 \), \( Q = 0.1 \), and \( \Theta = 0.01 \). The transition point separating stable from unstable configurations is identified at \( (1.49, 0) \). In the plot, the pink region denotes the domain of stability, while the purple region represents instability. A comparison with results from the photon sphere analysis confirms that the critical orbits considered in this work lie within the unstable regime.

\subsection{Shadow radius}
The shadow of a spherically symmetric black hole can be calculated by the following expression \cite{perlick2015influence,konoplya2019shadow}, 
\begin{equation}
	\begin{split}\label{shadow}
	&{R_{\text{sh}}}= \sqrt {\frac{g^{\scriptscriptstyle {(\Theta},\scriptscriptstyle{Q)}}_{\phi\phi}}{-g^{\scriptscriptstyle {(\Theta},\scriptscriptstyle{Q)}}_{tt}}}\bigg |_{r=r_{\text{ph}}} \\ 
    &=\Big[-\left(\Theta ^2 r_\text{ph}^4 \left(r_\text{ph}^2 \left(2 M^2-4 M r_\text{ph}+r_\text{ph}^2\right)+Q^2 r_\text{ph} (5 r_\text{ph}-8 M)+4 Q^4\right)\right./\\ 
    &\left.+4 r_\text{ph}^8 \left(r_\text{ph} (r_\text{ph}-2 M)+Q^2\right)\right)\left[\Theta ^2 \left(3 Q^2 r_\text{ph} (3 r_\text{ph}-10 M)+M r_\text{ph}^2 (11 M-4 r_\text{ph})+14 Q^4\right)\right.\\
    &\left.-2 r_\text{ph}^4 \left(r_\text{ph} (r_\text{ph}-2 M)+Q^2\right)\right]\left(r_\text{ph} (r_\text{ph}-2 M)+Q^2\right) \Big]^{1/2}
	\end{split}
\end{equation}

\begin{figure}[ht!]
	\centering
	\includegraphics[width=100mm]{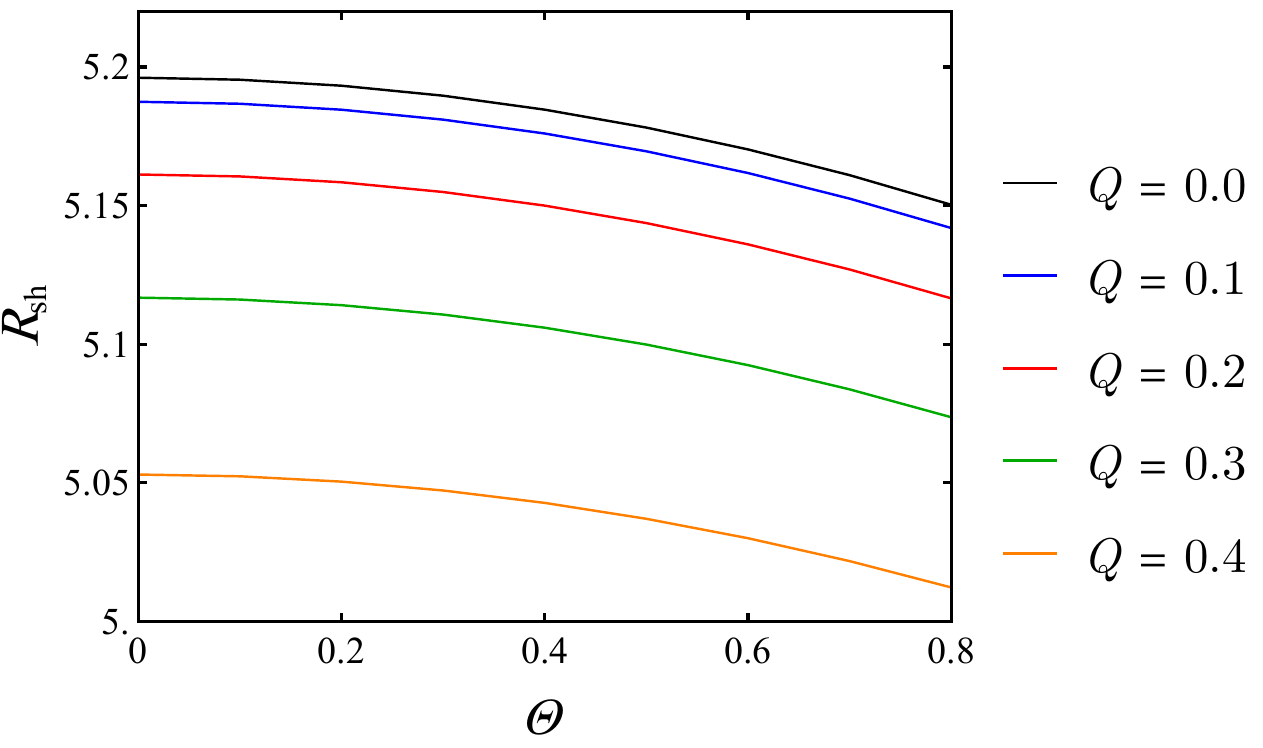}
	\caption{Shadow radius with respect to NC parameter $\Theta$ for $M = 1$ and various values of charge $Q$. }
	\label{fig:ShadowQ}
\end{figure}

In Fig. \ref{fig:ShadowQ}, we present an analysis of the shadow radius concerning the NC parameter for a few charge $Q$ for improved visualization.  

The shadow radius clearly decreases with an increase in the non--commutative parameter for all charge levels, indicating that $\Theta$ significantly influences the size of the black hole shadow. Conversely, given a constant NC parameter, a reduced charge results in an increased shadow radius.


\subsection{Gravitational lensing}

Building upon the expression for the Gaussian curvature obtained in Eq.~\eqref{gaussiancurvature}, the next step involves computing the deflection angle in the weak--field approximation by employing the Gauss--Bonnet theorem \cite{Gibbons:2008rj}. To facilitate this, the surface element on the equatorial plane is evaluated and can be written as
\ie
\mathrm{d}S = \sqrt{\gamma} \, \mathrm{d} r \mathrm{d}\varphi = \sqrt{\frac{g^{\scriptscriptstyle {(\Theta},\scriptscriptstyle{Q)}}_{rr}}{g^{\scriptscriptstyle {(\Theta},\scriptscriptstyle{Q)}}_{tt}}  \frac{g^{\scriptscriptstyle {(\Theta},\scriptscriptstyle{Q)}}_{\varphi\varphi}}{g^{\scriptscriptstyle {(\Theta},\scriptscriptstyle{Q)}}_{tt}} } \, \mathrm{d} r \mathrm{d}\varphi,
\fe
enabling us to calculate the deflection angle with the following expression

\ie
\begin{split}
& \hat{\alpha} (b,Q,\Theta) =  - \int \int_{D} \mathcal{K} \mathrm{d}S = - \int^{\pi}_{0} \int^{\infty}_{\frac{b}{\sin \varphi}} \mathcal{K} \mathrm{d}S \\
& \simeq  \, \frac{4 M}{b} -\frac{45 \pi  M^2 Q^2}{32 b^4}-\frac{8 M Q^2}{3 b^3}+\frac{3 \pi  M^2}{4 b^2}-\frac{3 \pi  Q^2}{4 b^2} -\frac{10415 \pi  \Theta ^2 M^2 Q^2}{1536 b^6} \\
& -\frac{1778 \Theta ^2 M Q^2}{75 b^5}+\frac{297 \pi  \Theta ^2 M^2}{128 b^4}+\frac{537 \pi  \Theta ^2 Q^2}{128 b^4}-\frac{16 \Theta ^2 M}{3 b^3}+\frac{\pi  \Theta ^2}{16 b^2}
\end{split}
\fe

\begin{figure}
    \centering
    \includegraphics[scale=0.51]{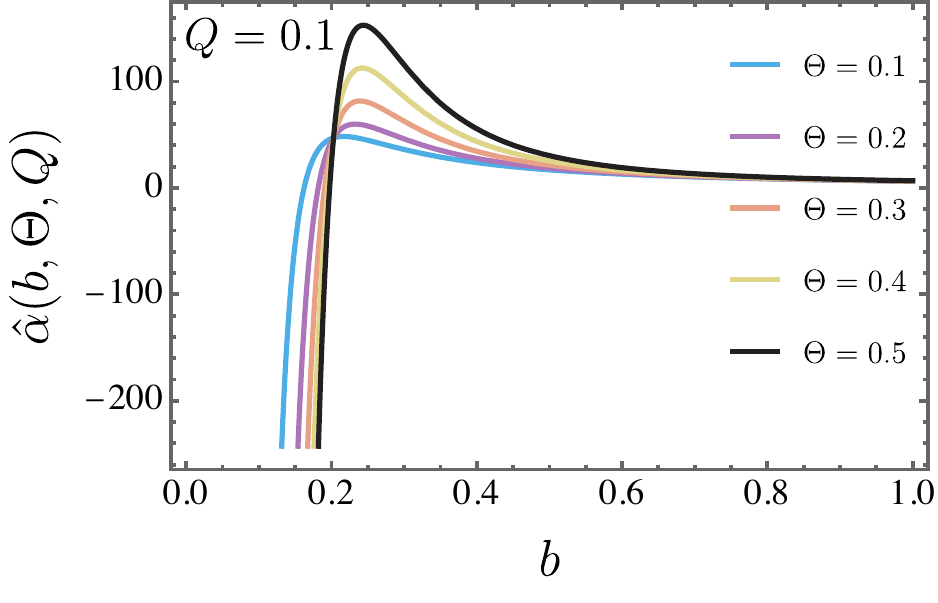}
    \includegraphics[scale=0.51]{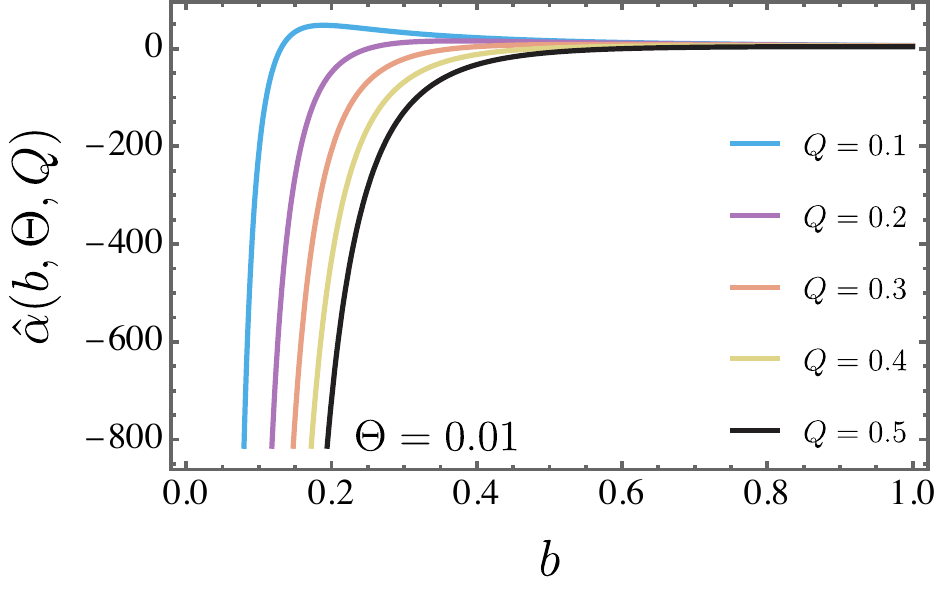}
    \caption{The deflection angle \( \hat{\alpha}(b, Q, \Theta) \) is plotted with respect to the impact parameter \( b \) for $M = 1$ and various values of the parameters \( \Theta \) and \( Q \).
}
    \label{angldefc}
\end{figure}

Figure~\ref{angldefc} depicts the variation of the deflection angle \( \hat{\alpha}(b, Q, \Theta) \). For a fixed impact parameter \( b = 0.3 \), an increase in the NC parameter \( \Theta \) corresponds to an enhancement in the magnitude of \( \hat{\alpha}(b, Q, \Theta) \). Conversely, raising the charge parameter \( Q \) produces a decrease in the deflection angle.

\subsection{Lensing observable}

By analyzing the black hole shadow observations of \(SgrA^* \) and \( M87 ^* \) obtained by the Event Horizon Telescope (EHT), we explore potential constraints on the NC parameter \( \Theta \) within the framework of a charged black hole in NC geometry.
 The angular diameter (radius) of the shadow, $d_\text{sh}$, plays a crucial role in probing deviations from classical general relativity. By comparing the theoretical shadow diameter predictions with EHT measurements, we assess whether the observational data places any bounds on the NC parameter.
A distant observer, located at a radial distance $\mathcal{D}$ from the black hole, measures the angular diameter $d_\text{sh}$ of the black hole shadow as  \cite{kumar2020rotating,afrin2023tests}
\ie
d_{\text{sh}} = 2 \frac{b_c}{\mathcal{D}}
\fe

\begin{figure}[ht!]
	\centering
	\includegraphics[height=58mm]{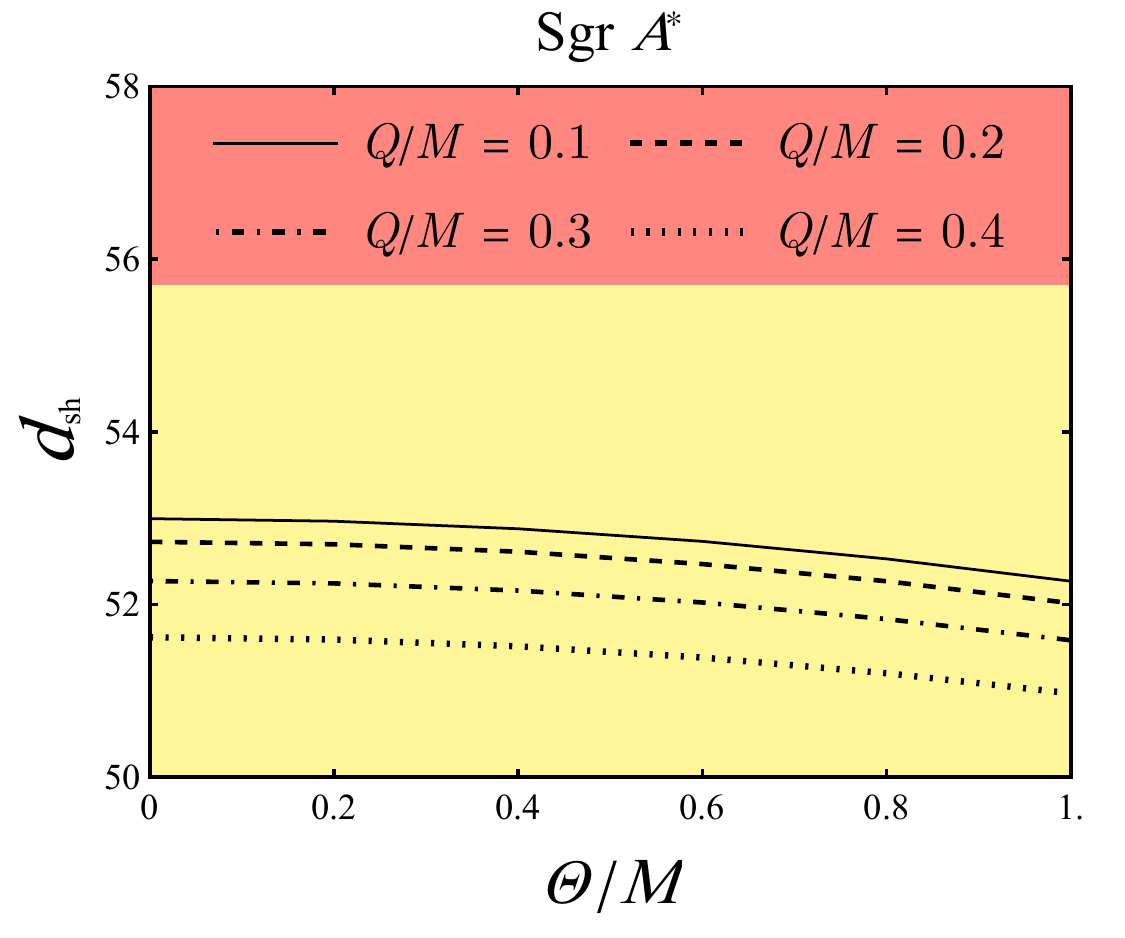}
    \includegraphics[height=58mm]{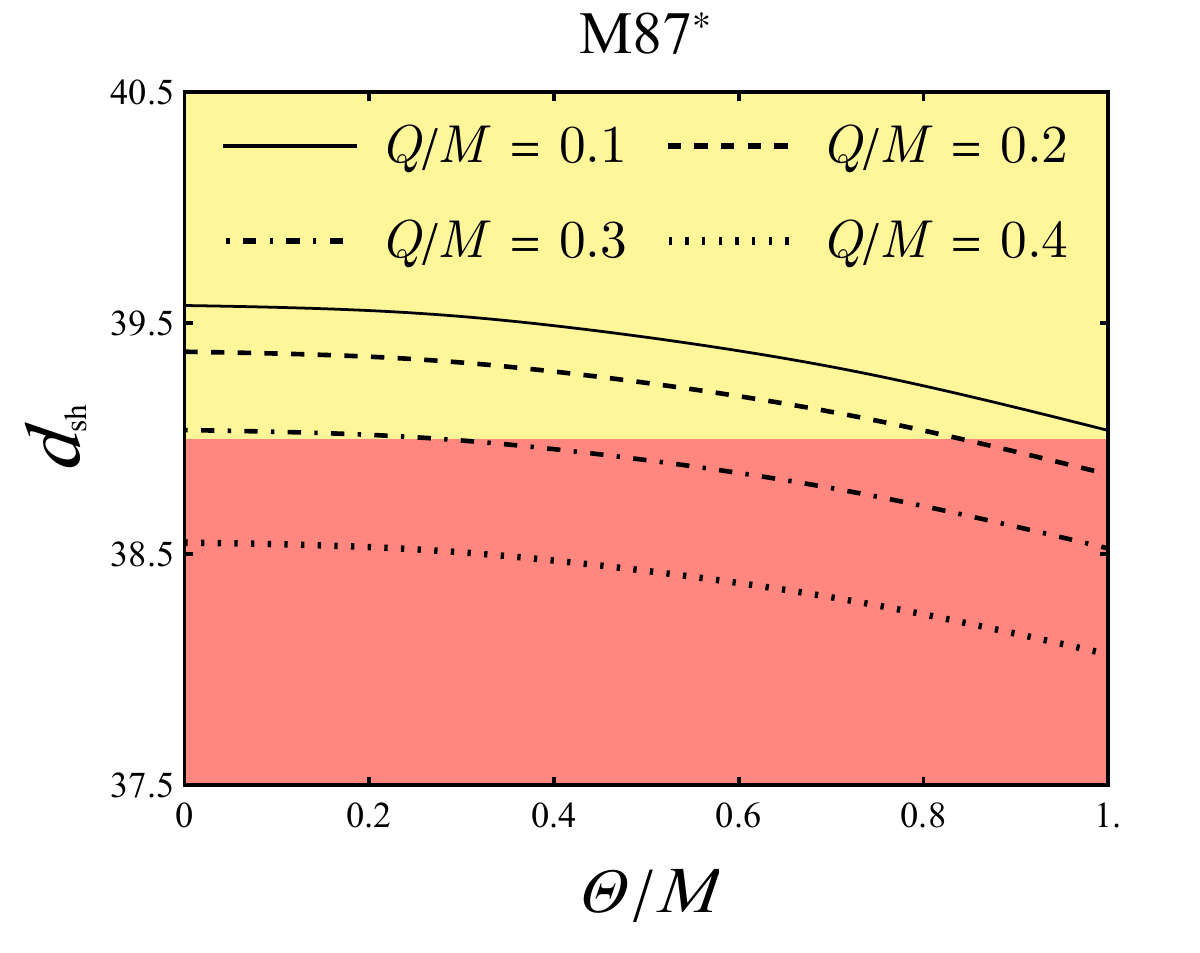}
	\caption{Angular diameter with respect to NC parameter $\Theta$ for based on $M$ and $\mathcal{D}$ of $Sgr A^*$ and $M87^*$. The yellow and orange areas represent the allowed and excluded regions of angular shadow, according to EHT observations.  }
	\label{fig:Cons}
\end{figure}

To explore the constraints on $\Theta$, based on both $Sgr A^*$ and $M87^*$ observations, the charged NC black hole shadow as a function of $\Theta$ is shown in Fig. \ref{fig:Cons}.

In the left panel, we assumed that the mass and distance to the observer were $M = 4  \times 10^6 M_{\odot}$ and $\mathcal{D} = 8.15 Kpc$, for the NC charged black hole as the $Sgr A^*$ black hole.

The angular diameter for different values of charge $Q$ and variation of NC parameter $\Theta$, does not show any constraint to be in the EHT collaboration reported mean value range $d_\text{sh}\in (41.5,55.7) \mu as$ \cite{akiyama2022firstSgr,akiyama2022firstSgrA}.

On the right panel, for the $M87^*$ case, we consider mass and distance to observer to be as $M = (6.5 \pm 0.7) \times 10^9 M_{\odot}$ and $\mathcal{D} = 16.8 Mpc$. The angular diameter of the $M87^*$ black hole shadow is inferred to be $d_{\text{sh}} = 42 \pm 3 \mu as$ based on the EHT data \cite{akiyama201987,akiyama2019M871}. In the context of the Reissner-Nordstr\"{o}m black hole as a model for $M87^*$, our analysis reveals specific constraints on the parameter \(\Theta/M\), which depend on the charge value \(Q/M\). Notably, when the ratio \(Q/M = 0.1\), no restrictions are placed on the NC parameter. However, at values of \(Q/M = 0.2\) and \(Q/M = 0.3\), the parameter \(\Theta/M\) is constrained to be less than $0.27$ and $0.83$, respectively. Furthermore, for \(Q/M = 0.4\), no permissible values for \(\Theta\) exist.


\section{Conclusion}

In this study, we have explored the profound implications of non--commutative geometry on the properties of a charged black hole, providing a detailed analysis of the thermodynamic behavior, quantum radiation, scalar perturbations, and optical characteristics. We have derived a consistent deformed spacetime geometry of a Reissner-Nordstr\"{o}m black hole that captures up to second--order corrections in the non--commutative parameter $\Theta$ within a four-dimensional, spherically symmetric (non-rotating) background, which naturally imposes certain limitations on the generality of our findings. 

The thermodynamic analysis revealed significant modifications to the black hole's thermal properties. The Hawking temperature and heat capacity were found to depend on both the charge $Q$ and the NC parameter $\Theta$, leading to the existence of a finite remnant mass at the end of the evaporation process. The study of Hawking radiation demonstrated distinct behaviors for bosonic and fermionic particles, with bosons exhibiting higher emission probabilities at low frequencies although the analysis does not include backreaction effects, which could further influence the evaporation process. 

Our investigation continued by examining the scalar perturbations. The calculated quasinormal modes showed increased oscillation frequencies and enhanced damping rates compared to the commutative case, with a significant imprint. Furthermore, non--commutativity breaks the degeneracy between different angular modes, offering a characteristic signature even though the parameter $\Theta$ itself remains phenomenological and lacks a well-established connection to Planck-scale physics. Moreover, the photon sphere and shadow analysis revealed that non--commutative correction results in a reduction in both the photon orbit radius and the apparent shadow size under the assumption of negligible metric backreaction. 

The gravitational lensing calculations in the weak--deflection limit demonstrated that both non--commutative geometry and the charge alter light deflection, suggesting that these effects might be detectable in future high--precision observations of gravitational lensing by compact objects. However, interpreting such potential observational signatures requires a clearer understanding of the fundamental origin of $\Theta$. 

Finally, the comparison of the lensing observables with Event Horizon Telescope (EHT) observations of $M87^*$ and $Sgr A^*$ has been discussed to obtain constraints on the NC parameter, although these bounds remain preliminary due to the simplifying assumptions of the model. 

The theoretical framework developed here may be applied as a foundation for further investigations into more complex scenarios, such as rotating black holes, potentially leading to new insights into the quantum nature of spacetime and helping address some of the present limitations by moving toward more realistic astrophysical configurations.

\section*{Acknowledgements} 

We would like to express our sincere gratitude to Professor A. A. Araujo Filho for his insightful discussions and valuable support with this project.

\appendix
\addcontentsline{toc}{section}{Appendices}
\renewcommand{\thesubsection}{\Alph{subsection}}

\subsection*{Appendix I}\label{AppA}
The coefficients of the effective potential in Eq. \eqref{Veff} are as follows
\begin{align*}
        &a^j_{lm}=\frac{m^2}{{\textit{N}_{lm}}} \int_{-1}^1 \frac{x^j P_l^m(x){}^2}{1-x^2} \, dx,\\
        &b^j_{lm}=\frac{1}{{\textit{N}_{lm}}}\int_{-1}^1 x^j P_l^m(x){}^2 \, dx,\\
        &c^j_{lm}=\frac{1}{{\textit{N}_{lm}}}\int_{-1}^1 x^j P_l^m(x) \left(\left(1-x^2\right) \frac{\partial ^2P_l^m(x)}{\partial x^2}-2 x \frac{\partial P_l^m(x)}{\partial x}\right) \, dx,\\
         &d^j_{lm}=\frac{1}{{\textit{N}_{lm}}}\int_{-1}^1 \left(1-x^2\right) \frac{\partial x^j}{\partial x} P_l^m(x) \frac{\partial P_l^m(x)}{\partial x} \, dx.
\end{align*}
\subsection*{Appendix II}\label{AppB}
The coefficients of the geodesic equations presented in Eq. \eqref{geot}-\eqref{geophi} are introduced as follows
\begin{align*}
    &\tilde{A}=r (r-2 M)+Q^2,\quad \tilde{B}=-75 \Theta ^2 M-2 r^3+18 \Theta ^2 r,\\
    &\tilde{C}=288 \Theta ^2 M^2+64 M r^3-254 \Theta ^2 M r-32 r^4+53 \Theta ^2 r^2,\\
    &\tilde{D}=-243 \Theta ^2 M-16 r^3+114 \Theta ^2 r,\quad \tilde{E}=8 \Theta ^2 M^2-32 M r^3-6 \Theta ^2 M r+16 r^4-\Theta ^2 r^2.\\
    &\tilde{F}=16 \Theta ^2 M^2+16 M r^3-15 \Theta ^2 M r-8 r^4+4 \Theta ^2 r^2, \quad \tilde{G}=-32 \Theta ^2 M+16 r^3+15 \Theta ^2 r.\\
    \end{align*}

\bibliography{main}
\bibliographystyle{ieeetr}

\end{document}